\documentclass[aps,pra, showpacs, floatfix,twocolumn,groupedaddress]{revtex4-1}

\usepackage{amsmath}

\usepackage{graphicx}
\usepackage[nolist]{acronym}
\renewcommand{\v}[1]{\ensuremath{\mathbf{#1}}} % for vectors

\newcommand{\abs}[1]{\left| #1 \right|} % for absolute value
\renewcommand{\d}[2]{\frac{d #1}{d #2}} % for derivatives
\newcommand{\dd}[2]{\frac{d^2 #1}{d #2^2}} % for double derivatives
\newcommand{\dn}[3]{\frac{d^{#1} #2}{d #3^{#1}}} % for nth derivatives
\newcommand{\pd}[2]{\frac{\partial #1}{\partial #2}}
\newcommand{\pdd}[2]{\frac{\partial^2 #1}{\partial #2^2}}
\newcommand{\ket}[1]{\left| #1 \right>} % for Dirac kets
\newcommand{\matrixel}[3]{\left< #1 \vphantom{#2#3} \right|
 #2 \left| #3 \vphantom{#1#2} \right>} % for Dirac matrix elements
\let\baraccent=\= % rename builtin command \= to \baraccent
\renewcommand{\=}[1]{\stackrel{#1}{=}} % for putting numbers above =
\providecommand{\e}[1]{\ensuremath{\times10^{#1}}}

\begin{document}

\preprint{}

\title{\acf{RMT} method for H$_2^+$ in short and intense laser fields}
% \title{\ac{RMT} method for describing H$_2^+$ in short and intense laser fields}


\author{Cathal \'O Broin}
%\email[]{cathal.obroin4@mail.dcu.ie}
\author{L. A. A. Nikolopoulos}
\affiliation{School of Physical Sciences, Dublin City University, Collins Ave, D9, Dublin, Ireland}
\affiliation{National Centre for Plasma Science and Technology, NCPST, Collins Ave, D9, Dublin, Ireland}
%\thanks{}
% \email[]{}
%\thanks{}


\date{\today}
\begin{abstract}
In this work we develop an approach for a molecular hydrogen ion (H$_2^+$) in the Born-Oppenheimer approximation while exposed to intense short-pulse radiation. Our starting point is the \acf{RMT} formulation for atomic hydrogen [L. A.A. Nikolopoulos \textit{et al}, Phys. Rev. A 78, 063420 (2008)] which has proven to be successful at treating multi-electron atomic systems efficiently and to high accuracy [LR Moore \textit{et al} J. of Mod. Opt. {\bf 58},1132, (2011)]. The present study on H$_2^+$ has been performed with a similar objective of developing an \textit{ab initio} method for solving the \acf{TDSE} for multi-electron diatomic molecules exposed to an external time-dependent potential field. The theoretical formulation is developed in detail for the molecular hydrogen ion where all the multi-electron and inter-nuclei complications are absent. As in the atomic case, the configuration space of the electron's coordinates are separated artificially over two regions; the inner ($I$) and outer ($II$) regions. In the region $I$ the time-dependent wavefunction is expanded on the eigenstate basis corresponding to the molecule's Hamiltonian augmented by Bloch operators, while in region $II$ a grid representation is used. We demonstrate the independence of our results on the introduced artificial boundary-surface by calculating observables that are directly accessed experimentally and also by showing gauge-dependent quantities are also invariant with the region $I$ box size. We also compare with other theoretical works and emphasize cases where basis-set approaches are currently very computationally expensive or intractable in terms of computational resources.
\end{abstract}

%\keywords{}

\maketitle

% \tableofcontents
\newpage
\section{Introduction}

Currently, a series of rapid developments are impacting strong field physics because of the discovery and refinement of a diverse range of radiation sources. The construction of \ac{FEL} sources capable of delivering unprecedented intense radiation in the soft and hard X-ray regimes has initiated new challenges, not only within atomic, molecular and optical physics but also for a number of other areas at the forefront of current interest more broadly, such as in biology, chemistry and nanoscience. Recently, another advancement on the opposite end of the wavelength regime, has been the increased availability of intense mid-IR range radiation \cite{Wolter2015}. In parallel with the aforementioned recent advances in strong-field physics the production of ultra-short pulses of sub-femtosecond duration has allowed the direct experimental observation of electronic and structural dynamics in matter \cite{Tzallas2011}. %\textbf{[Ref + myNaturePhysics]}.

These technological advances have gained significant attention from the theoretical community as the simulation of the experimental conditions requires significant computational resources which had not previously been feasible. For example, for fields into the infrared region, as pulse lengths increase and photon energies decrease, the number of angular momenta and the box sizes required in calculating photoelectron spectra increases. Thus, this regime is computationally very demanding and the problem becomes intractable, even for hydrogen (see \cite{Cormier1997,TetchouNganso2011}) and H$_2^+$ (see \cite{Kjeldsen2006}) which are the simplest atomic and molecular systems respectively.

It is a computationally difficult problem to treat the exact time-dependent response of a multielectron system subject to a strong \ac{EM} field by \textit{ab initio} methods. A number of theoretical groups, worldwide, are aiming for approaches beyond the computationally economical \ac{SAE} approximation \cite{Kamta2005,Palacios2005,Awasthi2005,Caillat2005,2006:29,Bandrauk2009,guan:2007,Guan2011,Moore2EtAl2011,Dundas2012,argenti:2013,feist:2014,Catoire2014};
the \ac{SAE} approach is a mature theoretical method and has been well explored \cite{Muller1999a,Awasthi2008}.

An alternative \textit{ab-initio} approach capable of treating multi-electron systems is based on R-matrix theory
%The basic formulation of the R-matrix approach first appeared in the context of nuclear theory research during the Manhattan Project \cite{Wigner1947} and was then later
applied to atomic and molecular systems for providing structural information \cite{Burke1971, Burke2011, Madden2011}.
The key concept in a R-matrix formulation is the \textit{division-of-space} concept which consists in separating the electrons' configuration space in two regions, namely the \textit{inner} and the \textit{outer} region. In the inner region, the atomic/molecular structure of the multielectron states are calculated with all the interactions taken into account while in the outer region only a single-electron wavefunction is required to be calculated \cite{Burke1971}. This \textit{division-of-space} approach appears to be well suited to tackle the computational problem which becomes especially crucial for the case of a multi-electron target. As a matter of fact, the full power of the R-matrix formulation is gained in the case of a multielectron target.

Traditionally, R-matrix approaches did not consider time dynamics in the study of collision and photoionization processes \cite{Burke1975a,Burke1975b}. Within the last decade, in response to technological and experimental advances, variants of time-dependent implementations utilizing the R-matrix computational framework have appeared which have been applied to specific atomic systems, namely the time-dependent R-matrix (TDRM) method \cite{burke:1997,Hart2007}, the time-dependent B-splines R-matrix (TDBSR) method \cite{guan:2007,guan:2008} and the present \ac{RMT} method  \cite{Nikolopoulos2008, Moore2EtAl2011,LysaghtEtAl2011}. An earlier application of the R-matrix theory to multiphoton processes appeared in the form of a Floquet expansion (RMF) of the driven time-dependent wavefunction \cite{Burke1991}. The RMF approach, although it is capable of treating the field non-perturbatively, cannot be considered as fully following the \ac{TDSE} solution methodology since it is only suited to laser pulses containing many cycles.
% \textbf{(P.G. Burke 2012 book, J. Tennyson, Physics Reports, 491, 29,(2010))}

The formulation of the TDRM method was developed and first applied to a one-dimensional model by Burke and Burke \cite{burke:1997} and later generalized and successfully applied to neon and argon \cite{lysaght:2008,lysaght:2013}. The TDRM method generates the R-matrix eigenstates in the inner region using an extension of the R-matrix Belfast codes \cite{rmatrx1:1995} to include a B-splines expansion of the continuum spectrum. Then the system's time-dependent wavefunction is propagated using a second-order Cayley propagator while in the outer region the time-dependent propagation is based on an R-matrix propagator, employed for solving a system of coupled second-order differential equations \cite{burke:2011}.

In the TDBSR method, the R-matrix eigenstates are generated using an alternative implementation of R-matrix theory on a non-orthogonal basis approach (B-splines) \cite{bsr:2006} and were propagated with an Arnoldi-Lanczos algorithm \cite{guan2009}. However the current implementation of the TDBSR method performs the wavefunction propagation only in an (enlarged) inner region, ignoring the outer-region completely, and as such does not fully take advantage of the division-of-space formulation of the R-matrix method. Nevertheless the obtained results are also indicative of the powerful machinery of the R-matrix methods at describing multiphoton processes in complex systems.

The \ac{RMT} method for atomic systems was mainly developed for the possibility of combining a high-order time-propagator for the multielectron wavefunction expanded on the inner-region R-matrix eigenstates with a finite-difference  method for the grid representation of the single-electron wavefunction in the outer region.

The method was first formulated and applied to the ionization of hydrogen \cite{Nikolopoulos2008} and then later extended to include the single-ionization of multielectron systems \cite{Moore2EtAl2011,LysaghtEtAl2011}. This \textit{unique} combination of an eigenstate-basis method with the finite-differences technique of propagating the time-dependent wavefunction has been recently demonstrated the capability of describing double-electron outer-region wavefunctions by Wragg \textit{et al} \cite{vanderhart:2015}.

\begin{figure}[ht]
\centering
\includegraphics[width=240px]{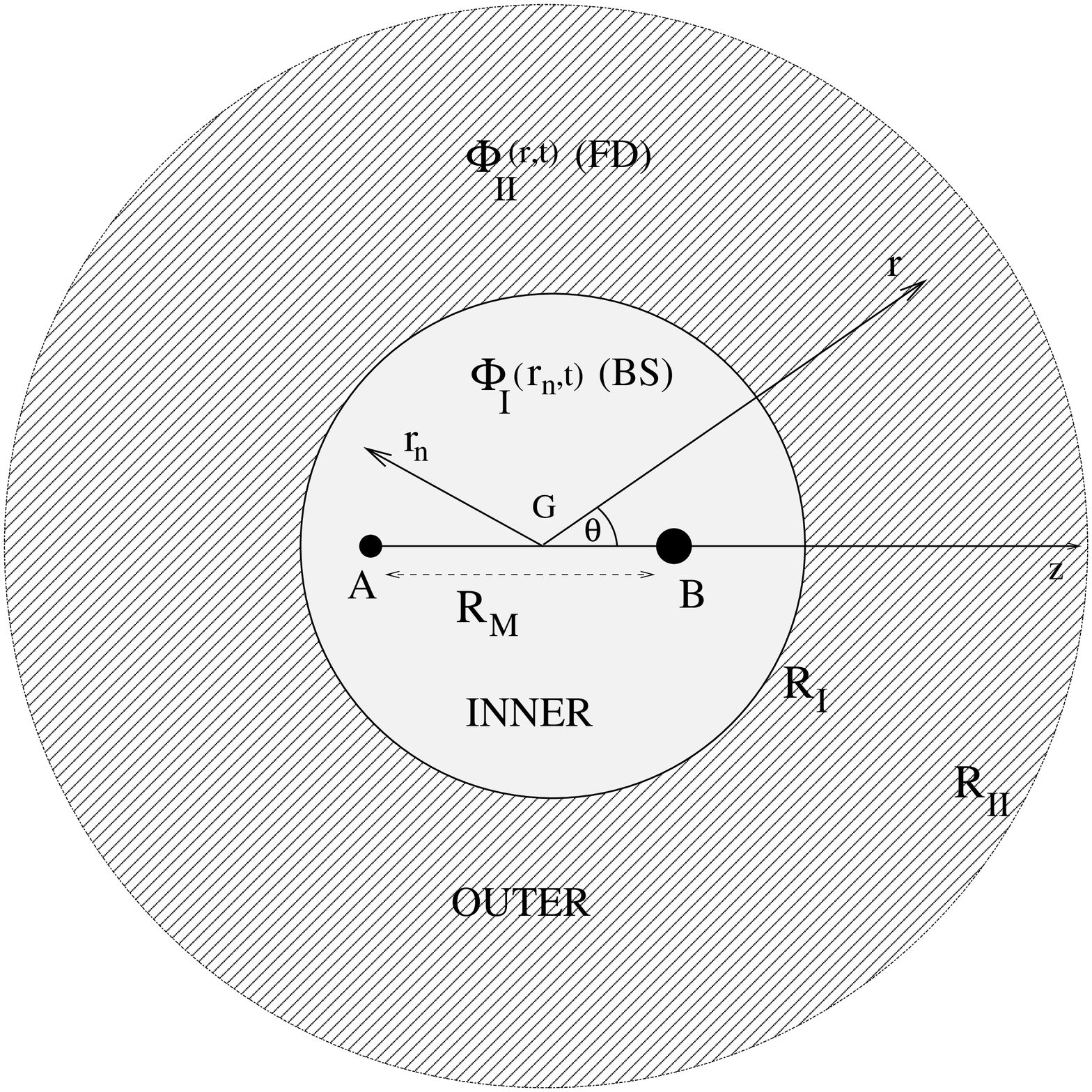}
\caption{Sketch of the \textit{division-of-space} method in the general diatomic situation (see \cite{Nikolopoulos2008} for a similar hydrogen diagram). The two nuclei are marked as A and B and the internuclear distance is $R_M$. With $\v r_n, 0 \le r_n \le R_I$ we denote collectively the positions of all electrons relative to a coordinate system with its origin placed at the molecule's center of mass, $G$, and the $z-$axis along the internuclear axis. 
With $\v r\equiv(r,\theta, \phi), R_I \le r \le R_{II}$ we represent the position of the ejected electron following the molecule's excitation/ionization by the external radiation.  
The molecule's configuration space is divided to two homocentric spherical boxes, (inner-region $I $ and outer-region $II$, with radii $R_I$ and $R_{II}$, respectively, with their center place at $G$. The molecule's time-dependent wavefunction in the inner region $I$ is expanded on an eigenstate basis (BS) $\Phi_I(\v r_n,t)$. In the outer region $II$ the system's wavefunction is described only by the ejected electron's wavefunction $\Phi(\v r,t)$ using a finite-difference (FD) approach. In the present case of one-electron system, $\v r_n \equiv \v r$.
\label{fig:RMatH2pPic}}
\end{figure}

The \ac{RMT} method allows the possibility of not only reducing the multi-electron dimensionality of the initial formulation to, effectively, a one (or two)-electron calculation in the (radially) larger region $II$ but also allows the use of different algorithmic approaches in the two different regions (Fig. \ref{fig:RMatH2pPic}). In region $I$ the power of the R-matrix method in calculating very accurate energy eigenfunctions and transition matrix elements is fully exploited. While, in the outer-region the extensibility of an equidistant, grid-based, representation of the wavefunction to very large distances, dramatically enhances the computational performance.

%That is, the \ac{RMT} approach has the advantage of the basis method over the finite difference of more accurate bound states, but also has the extensibility provided by grid methods.
%and the ability to switch to a hydrogenic representation of the molecular potential

%For example, the current outer region implementation supports arbitrary large extensions of the box-size without necessitating a recalculation of any basis elements but they do not allow for any major flexibility in the order of the difference operator or the grid spacing (it must be linear). The linear spacing is ideal for representing continuum states but does present difficulties for accurate representation close to the nuclei. The latter property has a large impact on the computational burden for both the purelly finite difference and the \ac{RMT} propagation methods, but not to a (purely) basis propagation approach. For an eigenstate basis, the switch-over has no impact on reducing the dimensionality, since the earlier coupling dictates the size of the Hamiltonian to be diagonalised by the TISE, while for the finite difference regions, the coupling instantly yields large computational and memory savings. This means the computational demand for arbitrarily large box sizes can be much lower in the \ac{RMT} case than for the basis approach.

In contrast with the above activities focused on atomic systems, analogous time-dependent formulations based on the division-of-space concept for molecular systems are not common. These methods include the t-surff method developed by the group of A. Scrinzi \cite{Tao2012, Scrinzi2012} and the analytical R-matrix method developed and applied in multielectron molecular systems \cite{Torlina2012a, Torlina2012b, Torlina2014}. Despite the fact that a number of groups worldwide have developed unique expertise in developing sophisticated ab initio methods and applied to solve the \ac{TDSE} for molecules it appears that these groups have not investigated the R-matrix method in this context.

In this work we introduce an approach based on an extension of the RMT method to H$_2^+$ without the complications arising from multi-electron considerations. Since the hydrogen molecule ion is also of interest for the \ac{RMT} approach as a stepping stone towards a full treatment of the hydrogen molecule and on to other polyatomic systems we develop the method in detail and demonstrate its applicability for diatomic one-electron systems with the EM field aligned along its symmetry-axis in the fixed-nuclei approximation.

In addition to the above scope, the present work also aims to ensure efficiency in the \textit{ab initio} description of molecules in intense and short \ac{EM} fields. To this end, we have developed the formulation (and implemented) for the velocity-gauge interaction operators in addition to the length-gauge alternative. As is generally known, for these studies, the velocity gauge is preferred against the length gauge for radiation in the long-wavelength regime because of its better convergence properties \cite{Cormier1996}. It is worth emphasizing here that the velocity-gauge approach implemented here has been employed for the first time within the time-dependent R-matrix approaches discussed above, and the existing \ac{RMT} codes. Moreover, the \ac{RMT} computations are done through \ac{GPGPU} techniques similar to our earlier works, albeit with some extra complexity \cite{OBroinNikolopoulos2012,OBroin2014}.

The paper is organized as follows. In Sec. ~\ref{sec:theory} we give an overview of the basic theory. It is the key section of this paper, where we set out in detail the theoretical formulation for the case of one-electron homo-nuclear diatomic molecules. The formulation presented in this section can be generalized to include the case of one-electron non-homonuclear diatomic systems with little extra effort. We also give the expressions for the calculation of experimental observables adapted to our methodology. In Sec. III to ensure the validity of the method, experimentally accessible quantities are compared against similar theoretical calculations available from the literature. Finally we have relegated the main technical details to the appendices. In the presentation of the formulas, atomic units are used ($m=\hbar=|e|=1$) throughout.

\section{Theoretical formulation} \label{sec:theory}
For the current case of short but intense fields, the Born-Oppenheimer approximation \cite{Bransden2003} is assumed so that the nuclei are effectively static over the short duration of the pulse. We also assume that the radiation field is linearly polarized along the symmetry axis of the diatomic molecule. Without loss of generality, we take the molecular axis as the $z-$axis of a $Oxyz$ Cartesian coordinate system. In this fixed-nuclei approximation the electronic Hamiltonian of the molecular hydrogen ion is given by
\begin{equation*}
H_0 = - \frac{\nabla^2}{2} - \frac{Z}{\abs{\v{r} - \v R_M/2}} - \frac{Z}{\abs{\v{r} + \v R_M/2 }}
\label{eq:h2p_0}
\end{equation*}
where $Z$ is the atomic number and $\pm \v R_M/2$ the position of the nuclei in the chosen coordinate system. Since the internuclear distance, $R_M=\abs{\v R_M}$, is treated as a constant term in the Hamiltonian, the $1/R_M$ term is omitted since it does not impact the electron dynamics.

The rotational properties of this one-electron diatomic system are more complex than the atomic single-electron case, thus making the problem considerably more demanding, both conceptually and computationally. Rotational symmetry is broken in H$_2^+$ since rotation of the system along the $x$ and $y$ axes is not equivalent to a rotation along the $z$ axis. This means the orbital angular momentum operator $\v L^2$ fails to commute with the Hamiltonian as $\v L^2$ is the generator of rotation, while the $L_z$ operator will still commute ($z-$axis projection). Thus the Hamiltonian and the orbital angular momentum operators do not have shared eigenfunctions. As a result, the time-dependent wavefunction of the system is not expanded in terms of a linear combination of mutual eigenfunctions of $H$ and $L_z$ and the parity operator $\Pi$ with associated eigenvalues $\epsilon$, $\mu$ and $\lambda$, respectively. The parity can be gerade (even, $\lambda=0$) or ungerade (odd, $\lambda=1$), reflecting whether or not the state is symmetric or anti-symmetric through a mirror reflection on the $x,y$ surface along the $z$ axis. In the particular case of study, the interaction of H$_2^+$ with a linearly-polarized field along the molecular axis, it is routinely shown that the excited states should have the same $\mu$ symmetry number as the initial state ($\mu = 0$). This allows us to neglect the $\mu$ quantum number, since the initial state in the present case is the H$_2^+$ ground state $1\sigma_g$.

The propagation of the electron wavefunction for a system of H$_2^+$ in the presence of an external laser field is calculated through the solution of the corresponding \ac{TDSE},
\begin{displaymath}
i\pd{}{t}\Phi(\v r, t) = \left[ H_0 + D(\v{r}, t) \right] \Phi(\v r, t).
\end{displaymath}
The expression for the basis expansion in terms of the energy eigenfunctions of $H_0$ is,
\begin{equation}
\Phi(\v r, t) = \sum_{n\lambda} C_{n\lambda}(t) \Phi_{n\lambda}(\v r),
\label{eq:1st_tdbs_wf}
\end{equation}
where $\Phi_{n\lambda}$ are solutions of the field-free Hamiltonian (see appendix A for the details) and the index $n$ is associated with the system's eigenenergies $(\epsilon_n \leftrightarrow n)$ and denotes both bound and (discretized) continuum eigenstates. Equivalently, the wavefunction can be also represented as a partial wave expansion:
\begin{equation}
\Phi(\v r, t) = \sum_{l} \frac{1}{r} f_l(r, t) Y_{l0}(\Omega),
\label{eq:tdfd_wf}
\end{equation}
where the eigenvalues of the angular momentum operator $\v L^2$ are characterized by the index $l$.  At this stage, further description of the calculational method requires the separate treatment of the \ac{TDSE} for the inner and the outer regions.

\subsection{TDSE in the inner region}
For the basis approach in region $I$ the \ac{TISE} associated with $H_0$ would be non-Hermitian if there was a naive division of the full space. To cope with the non-Hermiticity of the operators appearing in the \ac{TISE} the same approach as in the case of atomic hydrogen \cite{Nikolopoulos2008} is followed. First, the field-free Hamiltonian and the velocity gauge dipole-interaction operator are augmented to become their Hermitian counterparts within the spherical region $[0,R_I]$:
\begin{eqnarray}
\hat{H}_0(\v{r}) & = &  H_0(\v{r}) + \hat{ L}_h(\v{r}), \\
\hat{D}(\v{r},t) & = &  D(\v{r}, t) + \hat{L}_d(\v{r}, t),
\end{eqnarray}
where $\hat{L}_h$ and $\hat{L}_d$ are the corresponding Bloch operators (see appendix \ref{ap:Hermitise}). Then the augmented \ac{TISE} is diagonalised.

So for region $I$, the eigenfunction expansion in equation (\ref{eq:1st_tdbs_wf}) is modified to
\begin{equation}
\Phi_I(\v r, t) = \sum_{n\lambda} \tilde C_{n\lambda}(t) \tilde \Phi_{n\lambda}(\v r),
\label{eq:tdbs_wf}
\end{equation}
where $\tilde \Phi_{n\lambda}(\v r)$ are now energy eigenfunctions of the \textit{Bloch-augmented} Hamiltonian and $\tilde C_{n\lambda}(t)$ is the associated time-dependent coefficient.

Equating the expressions for the time-dependent wavefunction Eq. (\ref{eq:1st_tdbs_wf}) and Eq. (\ref{eq:tdfd_wf}), whilst decomposing the energy eigenfunctions in terms of spherical harmonics, we have
\begin{equation}
\sum_{n \lambda} \tilde C_{n \lambda}(t) \sum_{l \in l_\lambda} \frac{1}{r} P_{nl}(r)
Y_{l 0}(\theta) = \sum_{\lambda, l\in l_\lambda} \frac{1}{r} f_l(r, t) Y_{l 0}(\theta)
\end{equation}
and after some straightforward manipulations we arrive at the following time-dependent partial-wave relation in terms of the radial eigenstates $P_{nl_\lambda}$ for the inner region:
\begin{equation}
 f_{l_\lambda}(r, t) = \sum_{n} \tilde C_{n\lambda}(t) P_{n l_\lambda}(r), \quad  r \leq R_I. \label{eq:expandpartial}
\end{equation}
Clearly, this expansion only holds in the interval $(0 < r \le R_I)$ since the Bloch eigenfunctions are not defined outside this region.

Finally the initial \ac{TDSE} is maintained by subtracting both of the Bloch operators from the \ac{TDSE}. So the \ac{TDSE} expressed in terms of these Hermitian operators is
\begin{equation}
i \d{}{t} \Phi_I(\v r,t) = \left[ \hat{H}_0(\v r) + \hat{D}(\v r, t) + S(\v R_I,t) \right] \Phi_I(\v r, t)
\end{equation}
where $S(\v R_I,t) = - \hat{L}_h(R_I) - \hat{L}_d(\v R_I, t)$.
More specifically, in the current implementation, with the use of Eqns (\ref{eq:kinetic_bloch}),(\ref{eq:velocity}) the boundary term in the velocity gauge, is as below,
\[
S(\v R_I, t) = - \frac{1}{2}\delta(r-R_I) \left[ \d{}{r} + \frac{1}{r} 
 - i\frac{A(t)}{c}\cos \theta \right].
\]
This equation is fully equivalent to the initial TDSE without any approximation involved whatever. It is also worth emphasizing that the extra terms which are added and subtracted away are \textit{only non-zero on the boundary surface $r=R_I$}. The $\d{}{r}$ and $\frac{1}{r}$ terms in the RHS are due to the operator $\hat{L}_h$, while the last term is due to the interaction operator $\hat{L}_d$. The corresponding expression for the boundary term in the case of the length-gauge formulation is the same except that this latter term is not present.

In region $I$ the eigenfunctions and eigenvalues of the Hermitian Hamiltonian $\hat{H}_0$ should now be calculated. Then, the functions $\tilde \Phi_{n\lambda}(\v r)$ from Eq. (\ref{eq:tdbs_wf}) can be used to represent the inner-region portion of the \ac{TDSE} as a system of first-order ordinary differential equations. This will allow the calculation of the time-dependent coefficients, $\tilde{C}_{n\lambda}(t)$, at some time $t$ from a known initial state.

At this point we relegate the detailed development of the relevant formulation (namely, the calculational procedure for the field-free problem $\hat{H}_0 \Phi_{n\lambda}(r) = \epsilon_{n\lambda} \Phi_{n\lambda}(\v r) $ to the first two appendices.

Assuming that the field-free problem is solved, the wavefunction is expanded on these specific eigenfunctions of the field-free Bloch-augmented Hamiltonian to arrive at
\begin{widetext}
\begin{equation}
% i \d{}{t} \tilde{C}_{n\lambda}(t) = \sum_{n^\prime \lambda^\prime} \left [ \hat{H}_{n\lambda, n^\prime \lambda^\prime} + \hat{D}_{n\lambda, n^\prime \lambda^\prime}(t) \right ] \tilde{C}_{n^\prime \lambda^\prime}(t) + \matrixel{n\lambda}{S(\v R_I,t)}{\Phi_{II}(t)}, \label{eq:InnerRegionTDSE}
%Or without ket notation:
% i \d{}{t} \tilde{C}_{n\lambda}(t) = \sum_{n^\prime \lambda^\prime} \left [ \hat{H}_{n\lambda, n^\prime \lambda^\prime} + \hat{D}_{n\lambda, n^\prime \lambda^\prime}(t) \right ] \tilde{C}_{n^\prime \lambda^\prime}(t) + \int dr r^2 \int d\Omega \Phi_{n\lambda}(\v r) S(\v R_I, t) \Phi_{II}(\v r, t), \label{eq:InnerRegionTDSE}
%Or compactly:
i \d{}{t} \tilde{C}_{n\lambda}(t) = \sum_{n^\prime \lambda^\prime} \left [ \hat{H}_{n\lambda, n^\prime \lambda^\prime} + \hat{D}_{n\lambda, n^\prime \lambda^\prime}(t) \right ] \tilde{C}_{n^\prime \lambda^\prime}(t) + \int d^3 r \Phi_{n\lambda}(\v r) S(\v R_I,t) \Phi_{II}(\v r, t), \label{eq:InnerRegionTDSE}
\end{equation}
\end{widetext}
where $\hat{H}_{n\lambda, n^\prime \lambda^\prime}, \hat{D}_{n\lambda, n^\prime \lambda^\prime}(t)$ are the matrix elements of the operators $\hat{H}_0(\v r), \hat{D}(\v r, t)$, respectively. The time-dependent wavefunction $\Phi_{II}(\v r, t)$ is labeled with $II$ since we use the grid expansion (\ref{eq:tdfd_wf}). This expansion is also the one that is used in the outer-region $II$. Therefore, in the last term we do not use terms solely of the inner-region basis but rather $\Phi_{II}(\v R_I,t)$ since the $\delta(r-R_I)$ function in the boundary operator $S(t)$ contains derivatives. Calculation of the derivative of $\Phi_{II}(\v R_I,t)$ requires information from both region $I$ and $II$. The expansion of the inner region radial functions in terms of the partial waves through equation (\ref{eq:expandpartial}) at specific (inner-region) points allows the finite-difference spatial operators to be calculated by using the required values.

The reduction of the matrix element between a Bloch-energy eigenstate and angular momentum eigenstate implied by the coupling term of Eq. \ref{eq:InnerRegionTDSE} (final term on the RHS) is effectively the same as in the hydrogenic case \cite{Nikolopoulos2008}, except there is a summation over the partial-wave terms within the same symmetry. Summarizing all the above reductions we obtain the \ac{TDSE} in the inner region and in the velocity-gauge as,
\begin{widetext}
\begin{subequations}
\begin{eqnarray}
 i \d{}{t} \tilde C_{n \lambda}(t) &=&
      \epsilon_{n \lambda} \tilde C_{n \lambda}(t)
    + \sum_{n^\prime \lambda^\prime\neq\lambda} \tilde C_{n^\prime \lambda^\prime}(t) D_{n \lambda, n^\prime \lambda^\prime}(t) -\frac{1}{2} \sum_{l \in l_\lambda} P_{nl}(R_I) F_l(R_I,t),
    \label{eq:tdse_inner}
    \\
F_l(R_I,t) &=&\frac{d}{dr} f_l(R_I, t)
 - i \frac{A(t)}{c} \sum_{l^\prime = l \pm 1} K_{ll^\prime} f_{l^\prime}(R_I, t).
 \label{eq:tdse_inner_term}
\end{eqnarray}
\end{subequations}
\end{widetext}
In the length gauge the inner-region TDSE is obtained if we set $A(t) = 0$ in the above expression for the $F_l(R_I,t)$ term. The quantity $K_{ll^\prime}$, as given in  appendix \ref{ap:Eigenstates}, originates from the angular momentum properties of the dipole interaction term. Formally, the above inner-region TDSE differs with the corresponding one of the atomic case in that the appropriate quantum number to characterize stationary states changes from $l$ to $\lambda$ while the boundary surface term includes a summation over all the coupled (due to the molecular potential) orbital angular momenta $l$, belonging to the same symmetry $\lambda$.

\subsection{TDSE in the outer-region}
The molecular hydrogen ion grid form is derived in a standard way. The final equation in terms of a radial grid is
\begin{widetext}
\begin{eqnarray}
i\pd{}{t}f_l(r, t) &=&
\left[-\frac{1}{2}\pdd{}{r}
+\frac{l(l+1)}{2r^2}\right] f_{l}(r, t)
+ \sum_{l^\prime \in l_\lambda}
V_{l0, l^\prime 0}(r)
f_{l^\prime}(r, t) +\sum_{l^\prime \not\in l_\lambda} \hat{D}_{l 0,l^\prime 0}(r,t)f_{l^\prime}(r, t), \\
V_{l0, l^\prime 0}(r) &=& -2Z \delta_{(l + l^\prime) \text{ even}} \sqrt{(2l+1)(2l^\prime+1)} \sum_{L=\abs{l-l^\prime},\abs{l-l^\prime}+2,\dots}^{l+l^\prime} \frac{r_<^L}{r_>^{L+1}}
\begin{pmatrix} l & L & l^\prime \\ 0 & 0 & 0 \end{pmatrix}^2, \nonumber
\end{eqnarray}
\end{widetext}
where $r_> = \max(r,R_M/2)$ and $r_< = \min(r,R_M/2)$. The difference for the \ac{RMT} case is that the grid is not calculated over the full space but is limited to region $II$. The derivative for the point $r=R_I$ is calculated by using Eq. (\ref{eq:expandpartial}) to calculate the partial-wave grid points from within region $I$ as required. More details are shown in appendix \ref{ap:HigherDeriv}.

At large distances from the molecular center, relative to the separation of the nuclei ($r >> R_M$), the dominant part of the molecular potential is the spherically symmetric, being effectively hydrogenic. So, the potential term can be switched to that of a hydrogenic system to a very good approximation. In a regular basis calculation this wouldn't help to make the calculation more manageable in terms of size during the propagation because the symmetry of the system close to the nuclear center dictates the properties of the eigenfunctions.

\subsection{Bound-state populations and ionization yield.}
In the \ac{RMT} approach information from regions $I$ and $II$ are required to calculate observables and other quantities. In region $I$, the information must be extracted from the energy eigenfunctions $\tilde \Phi_{n\lambda}(\v r)$ and the associated coefficients $\tilde C_{n\lambda}(t)$. For region $II$, the partial waves $f_{lm}(r,t)$ and the spherical harmonics are available.

The most straightforward quantities to calculate for the basis, finite difference and \ac{RMT} methods for comparison, are the ground state population, the excited state population and the total ionization yield.

The ground state population $p_g(t)$ is calculated by the overlap of the initial state $\Phi(\v r,0)$ on to the state at time $t$, $\Phi(\v r, t)$. If the ground state of the system is initially populated, the evolution of the ground state population is,
\begin{equation}
p_g(t) = \abs{\int d^3r \Phi^\star(\v r, 0)\Phi(\v r, t)}^2
\label{eq:Generalp_g}
\end{equation}
Within the \ac{RMT} method the above integral is broken up into two separate integrals, one from $0$ to $R_I$ and another from $R_I$ to the outer boundary $R_{II}$:
\begin{multline}
p_g(t) = \bigg | \int_0^{R_I} dr r^2 \int d\Omega \, \Phi^\star(\v r, 0)\Phi(\v r, t) \\ + 
\int_{R_I}^{R_{II}} dr r^2 \int d\Omega \,\Phi^\star(\v r, 0)\Phi(\v r, t)  \bigg |^2,
\end{multline}
which is the statement that the ground state population has contributions from both regions $p_g(t) = \abs{p_{g;I}(t) + p_{g;II}(t)}^2,$ where $p_{g;I}(t)$ and $p_{g;II}(t)$ are the contributions from the respective regions.

For the region $I$ portion, the calculation is simply the sum
\begin{equation}
p_{g;I}(t) = \sum_{n\lambda} \tilde C^\star_{n\lambda}(0) \tilde C_{n\lambda}(t).
\label{eq:Basisp_g}
\end{equation}

In a basis calculation, the coefficient corresponding to $1\sigma_g$, $C_{10}(0)$, equals $1$ and the ground state population calculation is provided by the square of the corresponding time-dependent coefficient $\abs{C_{10}(t)}^2$, after the end of the pulse. Similarly the absolute value of the coefficient $\abs{C_{n\lambda}(t)}^2$, post-pulse, gives the population of the $\ket{n \lambda}$ eigenstate. 
In the \ac{RMT} method the absolute value of $\abs{\tilde C_{n\lambda}(t)}^2$ does not have the same relationship with the surviving populations of H$_2^+$. This because, in region $I$, the physical bound states are composed by a linear combination of all of the (normalized) RMT eigenfunctions within the symmetry. For this reason the ground state population requires the full calculation of the sum in equation (\ref{eq:Basisp_g}) rather than only the first term as in the basis method.
For the finite difference region, region $II$, an explicit spatial overlap must be calculated (approximately through the composite Simpson's rule) through direct integration:
\begin{equation}
p_{g;II}(t) = \sum_l \int_{R_I}^{R_{II}} dr f^\star_{l0}(r, 0) f_{l0}(r, t).
\label{eq:p_g}
\end{equation}

As a result, the total ground state population is
\begin{displaymath}
p_g(t) = \abs{\sum_{n\lambda} \tilde C^\star_{n\lambda}(0) \tilde C_{n\lambda}(t) + \sum_l \int_{R_I}^{R_{II}} dr f^\star_{l0}(r, 0) f_{l0}(r, t)}^2
\end{displaymath}

While the bound and ionized populations can also be trivially calculated by an appropriate summation in the basis case, for the \ac{RMT} case this information is lost (due to the R-matrix states mixing). Since the partial waves can be reconstructed, the approach for the finite difference case can also be used. For the finite difference case, the population of the bound states can be approximated by direct spatial integration of the probability values at all grid points inside a carefully chosen radius, say $r_i$:
\begin{equation}
p_b(t) = \sum_l \int_{0}^{r_i} dr |f_{l0} (r, t)|^2,
\label{eq:p_b}
\end{equation}
where care is taken in choosing $r_i$ such that the population of the bound state is converged with increasing radius.

Considering that the system only consists of bound and ionized states, knowing the bound state population means the ionized population is also known ($1-p_b(t)$). In the \ac{RMT} case, the exact equivalent of this bound population calculation can be done by simply getting the population of region $I$ through the coefficients, $\tilde C_{n\lambda}(t)$, (if $R_I$ is selected as the ionization boundary):

\begin{equation}
p_b(t) = \sum_{n\lambda} \abs{\tilde C_{n\lambda}(t)}^2.
\end{equation}

This also ensures that no numerical integration is required. Otherwise the same numerical integral as the ground state case can be used. It should be emphasized that the methods of calculating the ionization yield in the finite difference and \ac{RMT} methods both require post-pulse propagation. This means that the ionized population can not be easily distinguished from the bound state population during the run of the pulse. The, excited state population for example, is approximated by the quantity within a subset of the total box which is not counted towards ionization and which is not the ground state ($p_b(t) - p_g(t)$). Note that it is not until the post propagation, when the ionized population moves away from the central potential and the yield value asymptotically approaches a value, that the excited state population then becomes meaningful.

To ensure that a reasonable part of the contributions from continuum energy-eigenstates are counted in the ionization yield post-calculation, the wave equation is propagated forward in time. This allows the continuum contributions to move away from the central boundary so that the yield can be calculated by counting the probabilities after a certain cut-off radius ($R_I$ is ideal for the \ac{RMT} case). The radius should be chosen so that a minimal amount of bound state probability is included in the ionization yield calculation.

%%%%%%%%%%%%%%%%%%%%%%%%%%%%%%%%%%%%%%%%%%%%%%%%%%%%%%%%
\begin{figure}[!t]
\centering
\includegraphics[scale=0.3]{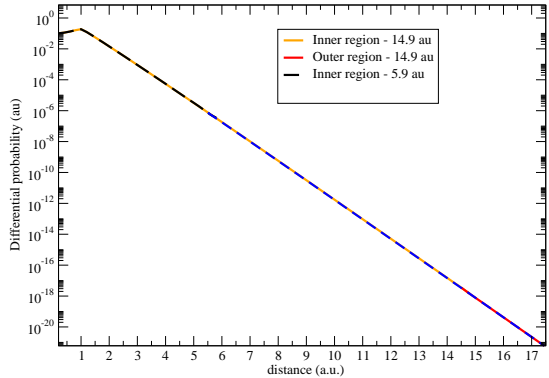}
\caption{A comparison of two H$_2^+$ \ac{RMT} diffusion calculations, with two different inner-outer region divisions; 5.9 a.u and 14.9 a.u. There are no major discrepancies between the different box sizes. The wavefunction calculated along $z$ is shown. }
\label{fig:GroundStateBSFDH2P}
\end{figure}
%%%%%%%%%%%%%%%%%%%%%%%%%%%%%%%%%%%%%%%%%%%%%%%%%%%%%%%

%%%%%%%%%%%%%%%%%%%%%%%%%%%%%%%%%%%%%%%%%% Fig 3
\begin{figure}[!t]
\centering
\includegraphics[scale=0.3]{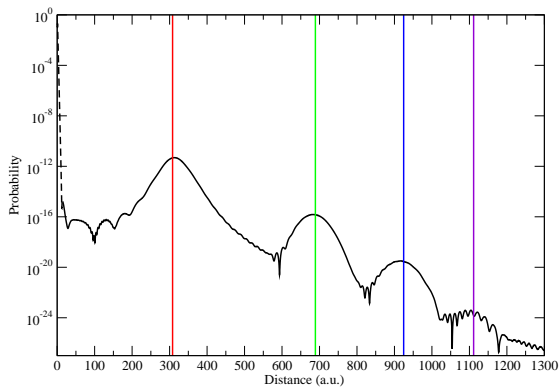}
\caption{Projection of the wavefunction along $z$ after a 24 cycle laser pulse and post-propagation 3 times the length of the pulse is shown. The inner region portion of the wavefunction is shown as a dotted black line, while the outer region is a solid black line. The expected positions of the different wavepacket peaks are also marked, showing 1 photon (red), 2-photon (green), 3-photon (blue) and 4-photon (violet) absorption electron wavepackets.}
\label{fig:10eV24cyclesw40ev}
\end{figure}
%%%%%%%%%%%%%%%%%%%%%%%%%%%%%%%%%%%%%%%% Fig 3

\section{Results and discussion} \label{Chapter:Compare}
Prior to discussing particular applications it is worth discussing computational issues at a more general level.  The eigenstate-basis method for the solution of the TDSE requires the precomputation of the associated field-free Hamiltonian eigenproblem (say of size $n_{max}$) for partial waves up to $l_{max}-1$. The latter numbers are dictated mainly by the strength and the duration of the EM field. Intense fields induce multiphoton absorptions that result to populating states with higher energy and orbital angular momentum quantum numbers.

For a reasonable energy spacing including higher energies ranges and to ensure the box is sufficiently large to capture the dynamics, a larger box size with a spatial discretisation which is sufficiently dense enough to represent the highest energies is required. Both effects increase the size of the eigenproblem. For example,
if the knot point spacing in the basis-grid is taken to be approximately equidistant over the full box size, then the computational effort for the number of eigenvalues and eigenvectors, $n_{max}$, to be found from the diagonalisation will scale linearly with both the box radius and also $l_{max}$. Further, the corresponding transition matrix element block scales with the square of both the box radius and $l_{max}$. Approximately commensurate with the increase in the number of states per partial wave, the number of partial waves included must also be increased to account for the greater occupancy of partial waves which have a higher angular momentum number.  This results in a rough scaling law for the total number of the states to be included of $\sim (n_{max} l_{max}/2)$ while the number of transition matrix elements required for computation will be $ \sim n_{max}^2 l_{max}^2/4$. In practice, both of these numbers are determined by testing the convergence of the specific value or set of values under study, such as the expectation value etc, against a gradual increase of the available parameters. Consequently, the total impact of increasing the size of these parameters is to make the field-free precomputation as well as the propagation of a basis-calculation prohibitively expensive. It leads to the highly undesirable condition of a Hamiltonian which should be approximated in a large spatial region with a very fine grid and a large number of coupled angular momentum terms.
It is for the above reasons that we have relied on the present extension RMT method combined with a parallel implementation for diatomic systems. The RMT approach only requires a basis representation in a restricted region of space, which thus restricts the computational burden significantly. Whatever computational hurdle remains, is then tackled with a parallel treatment. We now turn to the calculation of population and ionization yields of H$_2^+$ for specific cases.

%%%%%%%%%%%%%%%%%%%%%%%%%%%%%%%%%%%%%%%%%%%%%%%%%%%%%%%%%%%%% fig 4.
\begin{figure}[!t]
\centering
\includegraphics[scale=0.3]{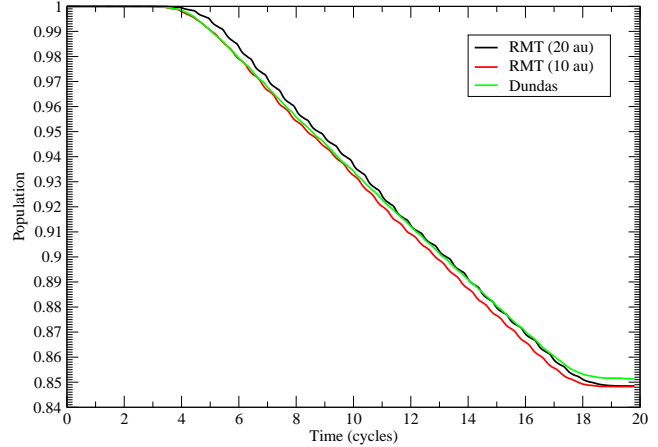}
\caption{The bound state population comparing against digitized data from Dundas et al \cite{Dundas2000}. The pulse is a trapezoidal pulse with a central photon energy of 5.4523 eV with a 4-cycle cosine ramp, a 12-cycle main portion and a 4-cycle cosine ramp down. The intensity is $4\e{14}$ Wcm$^{-2}$. }
\label{fig:Dundas}
\end{figure}
%%%%%%%%%%%%%%%%%%%%%%%%%%%%%%%%%%%%%%%%%%%%%%%%%%%%%%%%%%%%%
\subsection{Calculation of the initial state of H$_2^+$  by propagating the diffusion equation.}
The calculation of the initial state proceeds by propagation of the resulting (diffusion) equations in the inner and outer region after setting the external EM field equal to zero and by removing the complex number from the \ac{TDSE}.
Formally, the associated diffusion equation is obtained by making the substitution $t   \rightarrow \imath  t$ in the original field-free TDSE. In our calculation we set $l_{max} = 24$.
We compare our RMT calculations for the ground state of H$_2^+$ using a purely basis method \cite{OBroinNikolopoulos2012} (BS) and a purely finite-difference method (FD) after having extended the finite difference code in ref. \cite{OBroin2014} to include the H$_2^+$ system.  As previously reported by Martin \cite{Martin1999} 20 angular momentum terms provides a very decent estimate of the H$_2^+$ ground state energy ($-1.102 50$ a.u compared to an exact value of $1.102 63$ a.u). This means the 24 partial waves we have used should provide a reasonable measure. The calculated energy is $E^{FD}_g = -1.0942 $ a.u in the purely finite difference method, an error of about 0.765\%. In the present \ac{RMT} calculation, which consists of a mixed basis-finite difference propagation of the diffusion equation, the energy found is $E^{RMT}_g = -1.102532 $  a.u. This is very close to that achieved by Mart\'in \cite{Martin1999} (1.10250 a.u. for 20 $l_{max}$). We have also calculated the ground state eigenenergy with a purely eigenstate basis and found $E_g^{BS} = -1.102532$.
The outer-region box radius was $R_{II} = 713.1$ a.u. for the RMT and the FD method.
For the BS  method  $R_{II} = 99.9$ a.u.. For the RMT and the BS method the B-splines were of order $k=10$. In the RMT method the boundary radius was $R_I = 12.9$ a.u. The chosen grid spacing in the outer region was $dr = 0.2$ a.u. In the inner region, the chosen knot sequence $t=(t_1,t_2,..t_{n_b})$ of the B-splines grid was, $t$ =(0.0, 0.125, 0.25, 0.375, 0.5, 0.625, 0.750, 0.875, 0.95, 1.00, 1.05, 1.125, 1.25, 1.5, 1.7, 1.9, ... (increments of 0.2) ... 12.9) in atomic units.

In Fig. \ref{fig:GroundStateBSFDH2P} we show the wavefunctions evaluated along the $z-$axis with two different inner-outer region boundary sizes at $R_I=5.9$ a.u. and $R_I = 12.9$ a.u.. We see that the H$_2^+$ results are consistent regardless of box size. Note that a finer grid spacing along the boundary is required in the present RMT approach than would strictly be required in a standard basis calculation of the same size, since now the actual radial representation is important and not just the dipole and energy values.

\subsection{Populations and ionization yields H$_2^+ $ under EM fields.}
Being confident in the calculation of our ground state through the \ac{RMT} diffusion propagation method, we now look at the time-evolution of the wavefunction in an external \ac{EM} field.

First we can consider the case of making sure the dynamics follows what we would expect in terms of multiphoton absorption. We take a pulse that is so short that most of the ionization takes place at the peak of the pulse. If the initial wavefunction is treated as though it were initially localized in a small region close to $r=0$, which is indeed the case for the ground state, we then propagate the system such that the ionized portions of the wavefunction travel distances several times longer than the length of the molecule. We use a sine-squared pulse of carrier frequency $\omega = $ 40 eV so that any photon absorptions bring the H$_2^+$ straight into the continuum with a velocity corresponding to 10 eV for the case of one photon absorption. The different wavepackets corresponding to the different number of absorbed photons (\ac{ATI} peaks) should spatially separate out as they correspond to different acquired velocities. The formula for the distance away from the molecule is then quite simple in atomic units; $ r_n \sim (t-t_i) \sqrt{2E_n}$ a.u. with $n=1,2,..$.

In Fig. \ref{fig:10eV24cyclesw40ev} we plot a snapshot of the radial wavefunction. The wavepacket peaks are clearly separated and align with the expected distance considering the photon absorption count and the total ionized propagation time which is $ 3.5\tau_p$ if the total duration of the calculation is $t = 4 \tau_p$ (since the travel time of the wavepacket is after the time that the ionization takes place which we take this to be at the pulse's peak time at $t_i \sim 0.5 \tau_p$).

% Dundas 1
In Fig. \ref{fig:Dundas}, we show the results by Dundas et al \cite{Dundas2000} against the current \ac{RMT} calculation. In the figures we show the results by assuming two boundaries for the ionization thresholds for the calculation of bound state population;  $R_I = 10$ a.u and $R_I = 20$ a.u. In calculating the bound state population we assume the norm of the time-dependent wavefunction at distances $r < R_I$. The final ionization yield for Dundas et al is 0.851 (Dundas) and 0.848 (10 a.u, 20 a.u, respectively); a disagreement of 0.353\%. Within the present \ac{RMT} approach the bound states are well represented as the knot density can be increased close to the atomic nuclei without having a major impact on the overall number of splines.

%%%%%%%%%%%%%%%%%%%%%%%%%%%%%%%%%%%%%%%%%%%%%%%%%%%% text with Guan comparisons
\begin{table}%[H] add [H] placement to break table across pages
\begin{tabular}{c | c | c | c | c}
% \begin{tabular}{| c || l | l | l |}
\hline\hline
$I_0$& FE-DVR & BS &  \ac{RMT} & \% \\
\hline
$10^{12}$  &  $2.330\e{-6}$  &  $2.325\e{-6} $  &  $2.325\e{-6}$ & $-0.22\%$  \\
$10^{13}$  &  $2.330\e{-5}$  &  $2.326\e{-5} $  &  $2.325\e{-5}$ & $-0.22\%$  \\
$10^{14}$  &  $2.327\e{-4}$  &  $2.328\e{-4} $  &  $2.328\e{-4}$ & $+0.43\%$  \\
$10^{15}$  &  $2.304\e{-3}$  &  $2.352\e{-3} $  &  $2.351\e{-3}$ & $-2.04\%$  \\
\hline\hline
\end{tabular}
\caption{A comparison between the yields from results by Guan \textit{et al} (FE-DVR, prolate spheroidal coordinates) against the basis method (BS) and the \ac{RMT} approach, both in spherical coordinates. The sine-squared pulse used has a photon energy of $\omega = $ 40 eV and is 10 cycles in duration. The peak intensity, $I_0$, is varied from $10^{12}$ Wcm$^{-2}$ to $10^{15}$ Wcm$^{-2}$. The percentage difference in the \ac{RMT} method from the FE-DVR method is also shown.
\label{tb:ionised}}
\end{table}

%Centered version, revtex doesn't place this correctly
% \begin{widetext}
% \begin{center}
% \begin{table}%[H] add [H] placement to break table across pages
% \begin{tabular}{| c || l | l | l |}
% \hline\hline
% $I_0$ (W/cm$^{2}$) & Yield (FE-DVR) & Yield (BS) & Yield (\ac{RMT}) \\
% \hline
% $10^{12}$ & $2.330\e{-6}$ & $2.325\e{-6} (0.22\%)$ & $2.325\e{-6} (0.22\%)$ \\
% $10^{13}$ & $2.330\e{-5}$ & $2.326\e{-5} (0.17\%)$ & $2.325\e{-5} (0.22\%)$ \\
% $10^{14}$ & $2.327\e{-4}$ & $2.328\e{-4} (-0.43\%)$ & $2.328\e{-4} (-0.43\%)$\\
% $10^{15}$ & $2.304\e{-3}$ & $2.352\e{-3} (2.08\%)$ & $2.351\e{-3} (2.04\%)$ \\
% \hline\hline
% \end{tabular}
% \caption{A comparison between the yields from results by Guan \textit{et al} (FE-DVR, prolate spheroidal coordinates) against the basis method and the \ac{RMT} approach, both in spherical coordinates. The sine-squared pulse used has a photon energy of $\omega = $ 40 eV and is 10 cycles in duration. The intensity, $I_0$, is varied from $10^{12}$ W/cm$^{2}$ to $10^{15}$ W/cm$^{2}$.
% \label{tb:ionised}}
% \end{table}
% \end{center}
% \end{widetext}
%%%%%%%%%%%%%%%%%%%%%%%%%%%%%%%%%%%%%%%%%%%%%%%%%%%%%%%%%%%%

Next, in Table (\ref{tb:ionised}) we compare our ionization yields obtained with the pure basis (BS) and RMT methods against the results obtained by Guan \textit{et al} \cite{Guan2011}. The \ac{RMT} yields are calculated by treating the inner region population as the bound state population after one additional laser pulse length of field-free propagation. The bases are heavily modified in the inner-region and have a continuum spacing down to 0.4 a.u up to a radial distance of 138 a.u. while $l_{max} = 15$. The velocity gauge was used in all calculations. The pulse used has sine-squared envelope with photon energy of $\omega = $ 40 eV and its duration was 10 cycles in total. Our results are in agreement with those by  \textit{Guan et al} within 0.5\% except for the highest intensity which disagrees by 2\%. This represents a good agreement, particularly considering the very different methods used; Guan \textit{et al} use a finite-element-DVR technique and prolate spheroidal coordinates. Comparing the basis and \ac{RMT} methods themselves, the results are effectively identical. As a last comment on these results, note that the calculated ionization yields have linear dependence on the pulse's peak intensity $Y \sim I_0$, consistent with the fact that the ionization is possible by a single-photon absorption since $\omega = 40 $ eV.

%%%%%%%%%%%%%%%%%%%%%%%%%%%%%%%%%%%%%%%%%%%%%%%%%%%%%%%%%%%%%%
\begin{figure}[!t]
\centering
\includegraphics[scale=0.3]{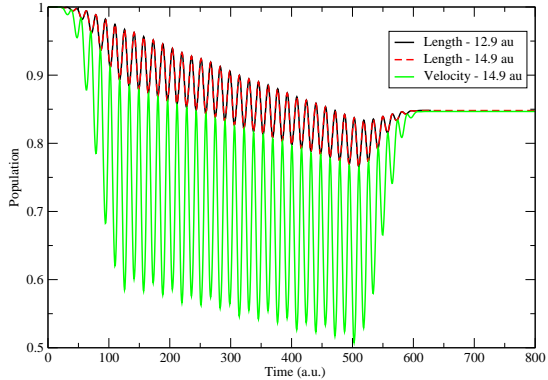}
\caption{Comparison of the ground state population as a function of time for H$_2^+$ in the length and velocity gauges for the pulse parameters given in \cite{Dundas2000}. Also shown is two inner-outer region boundary locations in the length gauge case; 12.9 a.u and 14.9 a.u. }
\label{fig:GroundPopBSFDH2P}
\end{figure}
%%%%%%%%%%%%%%%%%%%%%%%%%%%%%%%%%%%%%%%%%%%%%%%%%%%%%%%%%%%%%%

\subsection{Length and velocity-gauge calculations}
Our next results are concerned in investigating the accuracy of the different gauges, namely the length and the velocity gauges. For these calculations we compare results obtained by the pure basis (BS) and \ac{RMT} methods.

In Fig. \ref{fig:GroundPopBSFDH2P} we plot the ground state population during the propagation. The population of field-free states are gauge dependent in the presence of an external EM field but gauge invariant when the vector potential returns to zero. Consistent with this fact we observe an agreement for the population corresponding to the different gauges, at times where $A(t)$ is equal to zero. For minima of the field the \ac{RMT} length gauge result agrees with the velocity gauge calculation to a high degree.

%%%%%%%%%%%%%%%%%%%%%%%%%%%%%%%%%%%%%%%%%%%%%%%%%%%%%%%%%% fig 6
\begin{figure}[!t]
\centering
\includegraphics[scale=0.3]{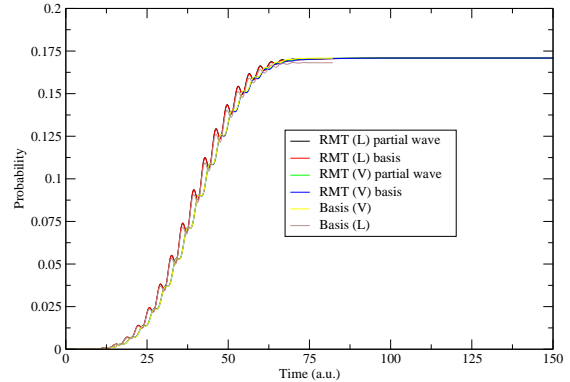}
\caption{The yield for the basis and \ac{RMT} methods in both the length and velocity gauges.}
\label{fig:Pulse_1_Yield_TDRM_BS}
\end{figure}
%%%%%%%%%%%%%%%%%%%%%%%%%%%%%%%%%%%%%%%%%%%%%%%%%%%%%%%%%%

In Fig. \ref{fig:Pulse_1_Yield_TDRM_BS} we plot the calculated ionization yields for the various methods and gauges. We see that there is clear agreement for the final yield
(as it should, since $A(t)=0$) and the yields during the pulse duration where $A(t)$ is equal to zero except for the basis length-gauge calculations.

\section{Conclusion}
From the development of the \ac{RMT} method in 2008 \cite{Nikolopoulos2008}, work since has focused on various aspects of atomic systems \cite{Moore2EtAl2011, LysaghtEtAl2011, vanderhart:2015}. In this paper, we have discussed the first extension of the \ac{RMT} approach to a molecular system. The work has focused on H$_2^+$ since it is the simplest molecular system. This extension reduces the dimensionality problems in H$_2^+$, since one can now have a full basis inner region and have a finite difference outer region which decouples the angular momenta terms as the hydrogenic approximation becomes valid ($V(r) \approx V_H(r)$ ).
The \ac{RMT} method has been expanded to include the case where eigenstates contain a mixture $l$ so that there is a transformation from an ungerade/gerade representation in the inner region to a representation consisting of a spherical-harmonic expansion. This work on H$_2^+$ has also been performed with the objective of approaching a treatment of H$_2$ which treats both electrons with full-correlation in an inner region but has one-electron outer region trajectories. It is hoped that future work will expand on this existing formulation and code base and extend the \ac{RMT} method to this new case.

\acknowledgments
The authors would like to acknowledge the contributions of Prof K.T. Taylor, Prof P. Decleva and Dr Daniel Dundas during the preparation of this work. KT and PD for their constant interest and stimulating discussions. In addition, PD provided technical data used for comparisons during development while DD kindly ran calculations for comparison. During the preparation of this work support was provided by the FP7 Grant ERG-HPCAMO/256601 project and the COST Actions \textit{XUV/X-ray light and fast ions for ultrafast chemistry} (CM1204) and \textit{Advanced X-ray spatial and temporal metrology} (MP1203).

\appendix

\section{H$_2^+$ Hamiltonian and dipole operator basis representation in a finite-region} \label{ap:Eigenstates}

Our starting point is the evaluation of the eigenstates of the electronic Hamiltonian for the molecular hydrogen ion, $ H \Phi_{n \lambda \mu}(\v{r}) = \epsilon \Phi_{\epsilon \lambda \mu}(\v{r})$ where $\epsilon$ is the eigenenergy value, $\lambda$ the parity symmetry (gerade/ungerade) and $\mu$ the projection of the angular momentum along the internuclear axis. Assuming a coordinate system with $z$-axis along the internuclear axis and with the origin placed in the middle of the nuclei's distance, for the case of $\mu = 0$ states
we can express the eigenstates on a spherical harmonic basis $Y_{lm_l}(\hat{r})$ as,
\begin{equation}
\Phi_{\epsilon \lambda}(\v{r}) = \sum_{l \in l_\lambda} \frac{1}{r} P_{\epsilon l}(r) Y_{l0}(\hat{r}).
\label{eq:wf_e}
\end{equation}

The $\sum_{l \in l_\lambda}$ indicates the summation over members of the set $l_\lambda$, where the sets are defined as:
\begin{align*}
 l_\lambda = \begin{cases}
  \left \{ 0, 2, 4, ..., l_{max}-2 \right \} & \forall\; \lambda = 0 \\
  \left \{ 1, 3, 5, ..., l_{max}-1 \right \} & \forall\; \lambda = 1. \\
 \end{cases}
\end{align*}
\noindent
The gerade/ungerade split of the angular momenta is imposed directly from analytic considerations of the molecular potential.
Projection on the spherical harmonic basis provides the corresponding Schr\"{o}dinger equation for the radial eigenfunctions $P_{nl}(r)$,
%\begin{widetext}
\begin{eqnarray*}
&&
\left[
 h_l(r) + \sum_{l^\prime \in l_\lambda} V_{l0, l^\prime 0}(r)
\right]  \frac{P_{\epsilon l^\prime}(r)}{r}
=
\epsilon_{n\lambda}\frac{P_{\epsilon l}(r)}{r} ,
\label{eq:tise_radial}
\\
&&  h_l(r) = -\frac{1}{2}\pdd{}{r} +\frac{l(l+1)}{2r^2}
%\right]\delta_{ll^\prime},
\label{eq:h0_radial}
\\
&&  V_{l0, l^\prime 0}(r) = -2Z\sqrt{2l+1}\sqrt{2l^\prime+1} \sum_L \frac{r_<^L}{r_>^{L+1}}
\begin{pmatrix} l & L & l^\prime \\ 0 & 0 & 0 \end{pmatrix}^2,
\label{eq:vm_radial}
\end{eqnarray*}
%\end{widetext}
with $L = \abs{l-l^\prime},\abs{l-l^\prime}+2,\dots, l+l^\prime$, and $l_> = \max(r,R_M/2), l_< = \min(r,R_M/2)$ and the bracket-like symbol being the 3j-symbol. For the H$_2^+$ case, the nuclear charge $Z$ is set to $1$.

In the general case, the eigenstates of the system are bound and continuum. The continuum wavefunctions are not square-integrable because they extend on infinitely and do not asymptotically approach zero with distance from the nucleus. Rather they are asymptotically periodic.
The first step to numerically calculate the eigenstates is to place inside a spherical box of, say, radius, $R_I$, \cite{Bachau2001, Awasthi2005, Barmaki2003}. This will discretize the full spectrum of the Hamiltonian (bound and continuum) and will make the continuum states square integrable. In this case, the eigenenergies can be characterized by a discrete index, as $\epsilon \rightarrow \epsilon_n \leftrightarrow n$, while their exact discretization will depend upon the $\lambda$ symmetry.

Here, we choose to expand the radial functions $P_{nl}(r)$ in terms of a non-orthogonal, local, polynomial set, namely the B-splines basis, $B_{ik}(r), i = 1,...,N_s$, as
\cite{Bachau2001,Martin1999}:
\begin{equation}
 P_{nl}(r) = \sum_i^{N_s} c_{i}^{(nl)} B_i(r).
\label{eq:pnl_bs}
\end{equation}
\subsection{H$_2^+$ Hamiltonian.}
Following on a standard procedure one can transform the TISE into a generalized eigensystem of the form:
\begin{equation}
 \sum_{l\in l_\lambda, l^\prime\in l_\lambda} \left ( \v H_l + \v V_{l^\prime, l} \right) \v C_l = \epsilon_{n\lambda} \sum_{l\in l_\lambda} \v S \v C_l , \label{eq:diag}
\end{equation}
where $\v S$ is the B-spline overlap matrix and the elements of the matrix are $ S_{ij} = \int_r dr B_i B_j$, $\v H_l$ is the B-spline overlap with the Hamiltonian,
\begin{equation}
 H_l^{(ij)} = \frac{1}{2} \int_0^{R_I} dr \left[ B^\prime_i B^\prime_j + l(l+1)\frac{B_iB_j}{r^2} \right]
 \end{equation}
and the elements of the molecular potential matrix $\v V_{l^\prime, l}$ given by $V_{l^\prime l}^{(ij)} = \int_0^{R_I} dr B_i V_{l0, l^\prime 0}(r) B_j$.

Since the matrices required can be explicitly calculated, the only unknowns are the specific eigenenergies ($\epsilon_{n \lambda}$) and the associated B-spline coefficients $c_i^{(nl)}$, gathered in vector $\v C_l$.

Thus, provided the matrices are symmetric (or the Hamiltonian representation Hermitian) the system can be diagonalised to produce real-valued eigenenergies. This is certainly the case if we impose the extra (boundary) condition on the possible solutions $P_{nl}(R_I) = 0$ at both ends (within the B-splines basis this easily done by excluding the first ($B_1$) and the last ($B_{n_s}$) B-splines
from the set in Eq. (\ref{eq:pnl_bs}). In the present work this is not the proper way to ensure Hermiticity (or symmetricity of the associated matrices) of the operators. The reason for this is that we require solutions which are non-zero on the boundary surface, namely solutions that $P_{nl}(R_I) \ne 0$ so that to ensure a non-zero probability current across the boundary surface. The
alternative way of 'Hermitizing' a non-Hermitian operator is by the addition of the so-called Bloch-operator, being another central concept in the R-matrix theory. We relegate a more detailed discussion on this Bloch-operator method in the next section to include any physical operator restricted in a finite region.

\subsection{Dipole operators}
%%%%%%%%%%%
The dipole operator in the length gauge is given by $D(\v{r},t) = \v{r} \cdot \v E(t)= E(t) r \cos \theta$, while for the velocity gauge it is $\hat{D}(\v{r}, t) = \v p \cdot \v{A}(t)/c$. After use of Eq. (\ref{eq:wf_e}) the corresponding matrix element equations are written as
\begin{eqnarray*}
\hat{D}_{n \lambda, n^\prime \lambda^\prime}(t) &=&  E(t)\matrixel{n\lambda}{r \cos\theta}{n^\prime, \lambda^\prime} \\
 &=& E(t)\sum_{l \in l_\lambda}\sum_{l^\prime \in l_{\lambda^\prime}} \matrixel{nl}{r \cos\theta}{n^\prime l^\prime},
\end{eqnarray*}
for the length-gauge dipole matrix elements, where use is made of the spherical harmonic expansion. For the velocity-gauge dipole matrix elements, the equivalent expression is
\begin{eqnarray*}
\hat{D}_{n \lambda, n^\prime \lambda^\prime}(t) &=&  \matrixel{n\lambda}{\v{p} \cdot \frac{\v A(t)}{c}}{n^\prime, \lambda^\prime} \\
 &=& \frac{A(t)}{c}\sum_{l \in l_\lambda}\sum_{l^\prime \in l_{\lambda^\prime}} \matrixel{nl}{-i\nabla \cdot \hat z }{n^\prime l^\prime},
\end{eqnarray*}

By the further use of the B-spline expansion of the radial solutions [Eq. (\ref{eq:pnl_bs})] and some angular momentum algebra one has the main expression in terms of known quantities:
\begin{equation}
D_{n\lambda,n^\prime \lambda^\prime}(t)= \sum_{l \in l_\lambda, l^\prime = l\pm 1} \frac{l_>}{ \sqrt{  4l^2_> - 1 } }
\mathcal G_{nl;n^\prime l^\prime}(t)\label{eq:dipole_r_1}
\end{equation}
with $l_> = max(l,l^\prime)$ and where $\mathcal G$ is the respective matrix, $\mathcal L$ or $\mathcal V$:
\begin{eqnarray*}
\mathcal{L}_{nl;nl^\prime}(t) &=& E(t) \int_{0}^{R_I} dr  P_{nl}(r) r P_{n^\prime l^\prime}(r),\label{eq:dipole_l_2} \\
\mathcal{V}_{nl;nl^\prime}(t) &=& \frac{A(t)}{c} \int_{0}^{R_I} dr P_{nl}(r) \left[\frac{d }{d r} + (l-l^\prime) \frac{l_>}{r}\right] P_{n^\prime l^\prime}(r),
\label{eq:dipole_v_2}
\end{eqnarray*}
where the specific choice depends on the gauge; length and velocity respectively. Thus, this can be calculated explicitly following diagonalisation of Eq. (\ref{eq:diag}).

\section{Restoring the Hermiticity of the field-free Hamiltonian operator in a restricted spatial region using the Bloch operator} \label{ap:Hermitise}

Now, we describe how the Bloch operator \cite{Bloch1957} is used to make the Hamiltonian Hermitian in the inner region $ 0 \le r \le R_I$ where the space of wavefunctions has arbitrary boundary conditions of the form $\alpha \Phi(\v R_I) + \beta \Phi(\v R_I) = 0$. The Laplacian operator will be Hermitian if \cite{QuantumMechanicsVol1Tannoudji},
\begin{equation}
 \matrixel{\Phi^\prime}{H}{\Phi} -  \matrixel{\Phi^\prime}{H}{\Phi}^\star = 0, \label{eq:Hermiticity}
\end{equation}
because the definition of a Hermitian operator is such that its matrix elements are equal to their own conjugate transpose.

This condition holds for all Hamiltonian terms that do not contain derivatives in the inner region. Applying Eq. (\ref{eq:Hermiticity}), the kinetic operator $T = - \nabla^2/2 $ term of the field-free part of the Hamiltonian and the dipole interaction term in the velocity gauge $D(\v r,t)= \frac{\v{A}(t)}{c} \cdot(-i \nabla) $ will be non-Hermitian operators. Both operators are now considered one at a time.

\subsection{Kinetic operator $ T$.} Substitution of Eq. (\ref{eq:wf_e}) in the formula (\ref{eq:Hermiticity}) followed by standard differential calculus manipulations results in
\begin{widetext}
\begin{equation}
 \matrixel{n\lambda}{ T}{n^\prime \lambda^\prime} - \matrixel{n^\prime \lambda^\prime}{ T}{n\lambda} = -\frac{1}{2}\sum_{l\in l_\lambda, l^\prime\in l_{\lambda^\prime}} \left[
 P_{nl}(R_I) \d{}{r} P_{n^\prime l^\prime}(R_I)  - P_{n^\prime l^\prime}(R_I) \d{}{r} P_{nl}(R_I)\right]
\label{eq:kinetic_hermite}
\end{equation}
\end{widetext}
where in order to ensure continuity at the center of symmetry the value $P_{nl}(0) = 0$ is used \cite{Martin1999}, while $P_{nl}(r)$ can take arbitrary values and arbitrary derivatives at the boundary.
At this stage we introduce the Bloch operator which is generally defined as \cite{Burke1971},
\begin{equation}
\hat{L}_h = \frac{1}{2}\delta(r-R_I) \left( \d{}{r} - \frac{\alpha-1}{r} \right),
\label{eq:kinetic_bloch}
\end{equation}
where $\alpha$ is a constant which can be chosen without constraints. As shown below, the following considerations are not affected by the particular choice of $\alpha$. In our calculations we have set $\alpha = 0$. For the Bloch operator $\hat{L}_h$ the following relation holds,
\begin{widetext}
\begin{equation}
 \matrixel{n\lambda}{\hat{L}_h}{n^\prime \lambda^\prime} - \matrixel{n^\prime \lambda^\prime}{\hat{L}_h}{n\lambda} = \frac{1}{2}
 \sum_{l\in l_\lambda, l^\prime\in l_{\lambda^\prime}}
 \left[
 P_{nl}(R_I) \d{}{r} P_{n^\prime l^\prime}(R_I)  - P_{n^\prime l^\prime}(R_I) \d{}{r} P_{nl}(R_I)
 \right].
\label{eq:block_hermite}
\end{equation}
\end{widetext}
Now by modifying the kinetic operator as $\hat{T} = T + \hat{L}_h$ and considering equations (\ref{eq:kinetic_hermite}) and (\ref{eq:block_hermite}), we find that:
\begin{equation}
 \matrixel{n\lambda}{\hat{T}}{n^\prime \lambda^\prime} - \matrixel{n^\prime \lambda^\prime}{\hat{T}}{n\lambda} = 0.
\end{equation}
Therefore, the matrix representation of the kinetic operator in the inner region, augmented by the above Bloch operator, is Hermitian and as such its diagonalization will provide real eigenvalues.

\subsection{Dipole interaction term in the velocity gauge.}

Following similar considerations as in the case of the kinetic operator we arrive at the result that an Hermitian velocity-gauge interaction operator can be obtained as,
$\hat{D} = D + L_d$, where $L_d$ is
\begin{equation}
L_d =  -i\frac{1}{2}\frac{A(t)}{c}\delta(r-R_I) \cos \theta.
\label{eq:velocity}
\end{equation}

\section{Recursive computation of the higher derivatives in the inner-region TDSE}
\label{ap:HigherDeriv}
Propagation of the inner-region TDSE Eq. (\ref{eq:tdse_inner}) through an explicit $p$-order propagator (as for example is the Taylor propagator) requires the evaluation $p$-times the Hamiltonian on to the wavefunction $\hat{H}^p \Phi(t)$. This operation will in turn result to the need to calculate the p-th order time derivative of the boundary surface term. In the below we give in more detail the relevant formulas for this calculation.

We start by noting the second derivative of the coefficients is given by the equation,
\begin{eqnarray*}
\dd{}{t} C_{n \lambda}(t)
 &=& -i \sum_{n^\prime \lambda^\prime} H_{n \lambda, n^\prime \lambda^\prime}(t) \d{}{t}C_{n^\prime \lambda^\prime}(t) \\
     &+&i \sum_{l \in l_\lambda} P_{nl}(R_I)  \d{}{t}F_l(r, t).
\end{eqnarray*}
The $p$-th derivative is
\begin{eqnarray*}
\dn{p}{}{t} C_{n \lambda}(t) &=& -i \sum_{n^\prime \lambda^\prime} H_{n \lambda, n^\prime \lambda^\prime}(t) \dn{p-1}{}{t}C_{n^\prime \lambda^\prime}(t) \\
&+& i \sum_{l \in l_\lambda} P_{nl}(R_I) \dn{p-1}{}{t}F_l(r, t).
\end{eqnarray*}

Analogous to the hydrogen case, from equation (\ref{eq:expandpartial}) the higher derivatives for the Taylor expansion on
a grid $r_j, j = 1,2,..$ are given by,
\begin{equation}
f_l^{(p)}(r_j, t) = - i \frac{dt}{p} \sum_{l^\prime} H_{l, l^\prime}(r_j, t) f^{(p-1)}_{l^\prime}(r_j, t)
\end{equation}
and
\begin{eqnarray*}
C^{(p)}_{n \lambda}(t) &=& \frac{-i dt}{p} \sum_{n^\prime \lambda^\prime} H_{n \lambda, n^\prime \lambda^\prime}(t) C^{(p-1)}_{n^\prime \lambda^\prime}(t) \\
&+& \frac{i dt}{p} \sum_{l \in l_\lambda} P_{nl}(R_I) F^{(p-1)}_l(r, t),
\end{eqnarray*}
where the partial wave terms on the inner boundary are recalculated as
\begin{equation}
f^{(p)}_{l_\lambda}(r_j, t) = \sum_{n} C^{(p)}_{n\lambda}(t) P_{n l_\lambda}(r_j) \;\;\;\; \forall \;\; r_j \leq R_I.
\end{equation}
The Taylor sums are then:
\begin{align}
& f_l(r_j, t + \tau) = \sum_p^P f_l^{(p)}(r_j, t) \quad \forall j \geq b,
\\& C_{n\lambda} (t + \tau) = \sum_p^P C^{(p)}_{n \lambda}(t).
\end{align}

%\begin{acknowledgments}
%\end{acknowledgments}

\bibliography{library}

\begin{thebibliography}{53}%
\makeatletter
\providecommand \@ifxundefined [1]{%
 \@ifx{#1\undefined}
}%
\providecommand \@ifnum [1]{%
 \ifnum #1\expandafter \@firstoftwo
 \else \expandafter \@secondoftwo
 \fi
}%
\providecommand \@ifx [1]{%
 \ifx #1\expandafter \@firstoftwo
 \else \expandafter \@secondoftwo
 \fi
}%
\providecommand \natexlab [1]{#1}%
\providecommand \enquote  [1]{``#1''}%
\providecommand \bibnamefont  [1]{#1}%
\providecommand \bibfnamefont [1]{#1}%
\providecommand \citenamefont [1]{#1}%
\providecommand \href@noop [0]{\@secondoftwo}%
\providecommand \href [0]{\begingroup \@sanitize@url \@href}%
\providecommand \@href[1]{\@@startlink{#1}\@@href}%
\providecommand \@@href[1]{\endgroup#1\@@endlink}%
\providecommand \@sanitize@url [0]{\catcode `\\12\catcode `\$12\catcode
  `\&12\catcode `\#12\catcode `\^12\catcode `\_12\catcode `\%12\relax}%
\providecommand \@@startlink[1]{}%
\providecommand \@@endlink[0]{}%
\providecommand \url  [0]{\begingroup\@sanitize@url \@url }%
\providecommand \@url [1]{\endgroup\@href {#1}{\urlprefix }}%
\providecommand \urlprefix  [0]{URL }%
\providecommand \Eprint [0]{\href }%
\providecommand \doibase [0]{http://dx.doi.org/}%
\providecommand \selectlanguage [0]{\@gobble}%
\providecommand \bibinfo  [0]{\@secondoftwo}%
\providecommand \bibfield  [0]{\@secondoftwo}%
\providecommand \translation [1]{[#1]}%
\providecommand \BibitemOpen [0]{}%
\providecommand \bibitemStop [0]{}%
\providecommand \bibitemNoStop [0]{.\EOS\space}%
\providecommand \EOS [0]{\spacefactor3000\relax}%
\providecommand \BibitemShut  [1]{\csname bibitem#1\endcsname}%
\let\auto@bib@innerbib\@empty
%</preamble>
\bibitem [{\citenamefont {Wolter}\ \emph {et~al.}(2015)\citenamefont {Wolter},
  \citenamefont {Pullen}, \citenamefont {Baudisch}, \citenamefont {Sclafani},
  \citenamefont {Hemmer}, \citenamefont {Senftleben}, \citenamefont
  {Schr{\"o}ter}, \citenamefont {Ullrich}, \citenamefont {Moshammer},\ and\
  \citenamefont {Biegert}}]{Wolter2015}%
  \BibitemOpen
  \bibfield  {author} {\bibinfo {author} {\bibfnamefont {B.}~\bibnamefont
  {Wolter}}, \bibinfo {author} {\bibfnamefont {M.~G.}\ \bibnamefont {Pullen}},
  \bibinfo {author} {\bibfnamefont {M.}~\bibnamefont {Baudisch}}, \bibinfo
  {author} {\bibfnamefont {M.}~\bibnamefont {Sclafani}}, \bibinfo {author}
  {\bibfnamefont {M.}~\bibnamefont {Hemmer}}, \bibinfo {author} {\bibfnamefont
  {A.}~\bibnamefont {Senftleben}}, \bibinfo {author} {\bibfnamefont {C.~D.}\
  \bibnamefont {Schr{\"o}ter}}, \bibinfo {author} {\bibfnamefont
  {J.}~\bibnamefont {Ullrich}}, \bibinfo {author} {\bibfnamefont
  {R.}~\bibnamefont {Moshammer}}, \ and\ \bibinfo {author} {\bibfnamefont
  {J.}~\bibnamefont {Biegert}},\ }\href@noop {} {\bibfield  {journal} {\bibinfo
   {journal} {Phys. Rev. X}\ }\textbf {\bibinfo {volume} {5}},\ \bibinfo
  {pages} {021034} (\bibinfo {year} {2015})}\BibitemShut {NoStop}%
\bibitem [{\citenamefont {Tzallas}\ \emph {et~al.}(2011)\citenamefont
  {Tzallas}, \citenamefont {Skantzakis}, \citenamefont {Nikolopoulos},
  \citenamefont {Tsakiris},\ and\ \citenamefont {Charalambidis}}]{Tzallas2011}%
  \BibitemOpen
  \bibfield  {author} {\bibinfo {author} {\bibfnamefont {P.}~\bibnamefont
  {Tzallas}}, \bibinfo {author} {\bibfnamefont {E.}~\bibnamefont {Skantzakis}},
  \bibinfo {author} {\bibfnamefont {L.~A.~A.}\ \bibnamefont {Nikolopoulos}},
  \bibinfo {author} {\bibfnamefont {G.~D.}\ \bibnamefont {Tsakiris}}, \ and\
  \bibinfo {author} {\bibfnamefont {D.}~\bibnamefont {Charalambidis}},\
  }\href@noop {} {\bibfield  {journal} {\bibinfo  {journal} {Nat. Phys.}\
  }\textbf {\bibinfo {volume} {7}},\ \bibinfo {pages} {781} (\bibinfo {year}
  {2011})}\BibitemShut {NoStop}%
\bibitem [{\citenamefont {Cormier}\ and\ \citenamefont
  {Lambropoulos}(1997)}]{Cormier1997}%
  \BibitemOpen
  \bibfield  {author} {\bibinfo {author} {\bibfnamefont {E.}~\bibnamefont
  {Cormier}}\ and\ \bibinfo {author} {\bibfnamefont {P.}~\bibnamefont
  {Lambropoulos}},\ }\href@noop {} {\bibfield  {journal} {\bibinfo  {journal}
  {J. Phys. B}\ }\textbf {\bibinfo {volume} {77}} (\bibinfo {year}
  {1997})}\BibitemShut {NoStop}%
\bibitem [{\citenamefont {{Tetchou Nganso}}\ \emph {et~al.}(2011)\citenamefont
  {{Tetchou Nganso}}, \citenamefont {Popov}, \citenamefont {Piraux},
  \citenamefont {Madro\~{n}ero},\ and\ \citenamefont
  {Njock}}]{TetchouNganso2011}%
  \BibitemOpen
  \bibfield  {author} {\bibinfo {author} {\bibfnamefont {H.~M.}\ \bibnamefont
  {{Tetchou Nganso}}}, \bibinfo {author} {\bibfnamefont {Y.~V.}\ \bibnamefont
  {Popov}}, \bibinfo {author} {\bibfnamefont {B.}~\bibnamefont {Piraux}},
  \bibinfo {author} {\bibfnamefont {J.}~\bibnamefont {Madro\~{n}ero}}, \ and\
  \bibinfo {author} {\bibfnamefont {M.~G.~K.}\ \bibnamefont {Njock}},\ }\href
  {\doibase 10.1103/PhysRevA.83.013401} {\bibfield  {journal} {\bibinfo
  {journal} {Phys. Rev. A}\ }\textbf {\bibinfo {volume} {83}},\ \bibinfo
  {pages} {013401} (\bibinfo {year} {2011})}\BibitemShut {NoStop}%
\bibitem [{\citenamefont {Kjeldsen}\ \emph {et~al.}(2006)\citenamefont
  {Kjeldsen}, \citenamefont {Madsen},\ and\ \citenamefont
  {Hansen}}]{Kjeldsen2006}%
  \BibitemOpen
  \bibfield  {author} {\bibinfo {author} {\bibfnamefont {T.}~\bibnamefont
  {Kjeldsen}}, \bibinfo {author} {\bibfnamefont {L.}~\bibnamefont {Madsen}}, \
  and\ \bibinfo {author} {\bibfnamefont {J.}~\bibnamefont {Hansen}},\ }\href
  {\doibase 10.1103/PhysRevA.74.035402} {\bibfield  {journal} {\bibinfo
  {journal} {Phys. Rev. A}\ }\textbf {\bibinfo {volume} {74}},\ \bibinfo
  {pages} {035402 (1} (\bibinfo {year} {2006})}\BibitemShut {NoStop}%
\bibitem [{\citenamefont {Kamta}\ and\ \citenamefont
  {Bandrauk}(2005)}]{Kamta2005}%
  \BibitemOpen
  \bibfield  {author} {\bibinfo {author} {\bibfnamefont {G.}~\bibnamefont
  {Kamta}}\ and\ \bibinfo {author} {\bibfnamefont {A.}~\bibnamefont
  {Bandrauk}},\ }\href {\doibase 10.1103/PhysRevA.71.053407} {\bibfield
  {journal} {\bibinfo  {journal} {Phys. Rev. A}\ }\textbf {\bibinfo {volume}
  {71}},\ \bibinfo {pages} {053407} (\bibinfo {year} {2005})}\BibitemShut
  {NoStop}%
\bibitem [{\citenamefont {Palacios}\ \emph {et~al.}(2005)\citenamefont
  {Palacios}, \citenamefont {Barmaki}, \citenamefont {Bachau},\ and\
  \citenamefont {Mart\'in}}]{Palacios2005}%
  \BibitemOpen
  \bibfield  {author} {\bibinfo {author} {\bibfnamefont {A.}~\bibnamefont
  {Palacios}}, \bibinfo {author} {\bibfnamefont {S.}~\bibnamefont {Barmaki}},
  \bibinfo {author} {\bibfnamefont {H.}~\bibnamefont {Bachau}}, \ and\ \bibinfo
  {author} {\bibfnamefont {F.}~\bibnamefont {Mart\'in}},\ }\href {\doibase
  10.1103/PhysRevA.71.063405} {\bibfield  {journal} {\bibinfo  {journal} {Phys.
  Rev. A}\ }\textbf {\bibinfo {volume} {71}},\ \bibinfo {pages} {63405}
  (\bibinfo {year} {2005})}\BibitemShut {NoStop}%
\bibitem [{\citenamefont {Awasthi}\ \emph {et~al.}(2005)\citenamefont
  {Awasthi}, \citenamefont {Vanne},\ and\ \citenamefont {Saenz}}]{Awasthi2005}%
  \BibitemOpen
  \bibfield  {author} {\bibinfo {author} {\bibfnamefont {M.}~\bibnamefont
  {Awasthi}}, \bibinfo {author} {\bibfnamefont {Y.~V.}\ \bibnamefont {Vanne}},
  \ and\ \bibinfo {author} {\bibfnamefont {A.}~\bibnamefont {Saenz}},\ }\href
  {http://stacks.iop.org/0953-4075/38/i=22/a=005} {\bibfield  {journal}
  {\bibinfo  {journal} {J. Phys. B}\ }\textbf {\bibinfo {volume} {38}},\
  \bibinfo {pages} {3973} (\bibinfo {year} {2005})}\BibitemShut {NoStop}%
\bibitem [{\citenamefont {Caillat}\ \emph {et~al.}(2005)\citenamefont
  {Caillat}, \citenamefont {Zanghellini}, \citenamefont {Kitzler},
  \citenamefont {Koch}, \citenamefont {Kreuzer},\ and\ \citenamefont
  {Scrinzi}}]{Caillat2005}%
  \BibitemOpen
  \bibfield  {author} {\bibinfo {author} {\bibfnamefont {J.}~\bibnamefont
  {Caillat}}, \bibinfo {author} {\bibfnamefont {J.}~\bibnamefont
  {Zanghellini}}, \bibinfo {author} {\bibfnamefont {M.}~\bibnamefont
  {Kitzler}}, \bibinfo {author} {\bibfnamefont {O.}~\bibnamefont {Koch}},
  \bibinfo {author} {\bibfnamefont {W.}~\bibnamefont {Kreuzer}}, \ and\
  \bibinfo {author} {\bibfnamefont {A.}~\bibnamefont {Scrinzi}},\ }\href@noop
  {} {\bibfield  {journal} {\bibinfo  {journal} {Phys. Rev. A}\ }\textbf
  {\bibinfo {volume} {71}},\ \bibinfo {pages} {12712} (\bibinfo {year}
  {2005})}\BibitemShut {NoStop}%
\bibitem [{\citenamefont {Nikolopoulos}\ and\ \citenamefont
  {Lambropoulos}(2006)}]{2006:29}%
  \BibitemOpen
  \bibfield  {author} {\bibinfo {author} {\bibfnamefont {L.~A.~A.}\
  \bibnamefont {Nikolopoulos}}\ and\ \bibinfo {author} {\bibfnamefont
  {P.}~\bibnamefont {Lambropoulos}},\ }\href@noop {} {\bibfield  {journal}
  {\bibinfo  {journal} {J. Phys. B}\ }\textbf {\bibinfo {volume} {39}},\
  \bibinfo {pages} {883} (\bibinfo {year} {2006})}\BibitemShut {NoStop}%
\bibitem [{\citenamefont {Bandrauk}\ \emph {et~al.}(2009)\citenamefont
  {Bandrauk}, \citenamefont {Chelkowski}, \citenamefont {Diestler},
  \citenamefont {Manz},\ and\ \citenamefont {Yuan}}]{Bandrauk2009}%
  \BibitemOpen
  \bibfield  {author} {\bibinfo {author} {\bibfnamefont {A.~D.}\ \bibnamefont
  {Bandrauk}}, \bibinfo {author} {\bibfnamefont {S.}~\bibnamefont
  {Chelkowski}}, \bibinfo {author} {\bibfnamefont {D.~J.}\ \bibnamefont
  {Diestler}}, \bibinfo {author} {\bibfnamefont {J.}~\bibnamefont {Manz}}, \
  and\ \bibinfo {author} {\bibfnamefont {K.-J.~J.}\ \bibnamefont {Yuan}},\
  }\href {\doibase 10.1103/PhysRevA.79.023403} {\bibfield  {journal} {\bibinfo
  {journal} {Phys. Rev. A}\ }\textbf {\bibinfo {volume} {79}},\ \bibinfo
  {pages} {23403} (\bibinfo {year} {2009})}\BibitemShut {NoStop}%
\bibitem [{\citenamefont {Guan}\ \emph {et~al.}(2007)\citenamefont {Guan},
  \citenamefont {Zatsarinny}, \citenamefont {Bartschat}, \citenamefont
  {Schneider}, \citenamefont {Feist},\ and\ \citenamefont {Noble}}]{guan:2007}%
  \BibitemOpen
  \bibfield  {author} {\bibinfo {author} {\bibfnamefont {X.}~\bibnamefont
  {Guan}}, \bibinfo {author} {\bibfnamefont {O.}~\bibnamefont {Zatsarinny}},
  \bibinfo {author} {\bibfnamefont {K.}~\bibnamefont {Bartschat}}, \bibinfo
  {author} {\bibfnamefont {B.~I.}\ \bibnamefont {Schneider}}, \bibinfo {author}
  {\bibfnamefont {J.}~\bibnamefont {Feist}}, \ and\ \bibinfo {author}
  {\bibfnamefont {C.~J.}\ \bibnamefont {Noble}},\ }\href {\doibase
  10.1103/PhysRevA.76.053411} {\bibfield  {journal} {\bibinfo  {journal} {Phys.
  Rev. A}\ }\textbf {\bibinfo {volume} {76}},\ \bibinfo {pages} {053411}
  (\bibinfo {year} {2007})}\BibitemShut {NoStop}%
\bibitem [{\citenamefont {Guan}\ \emph {et~al.}(2011)\citenamefont {Guan},
  \citenamefont {Secor}, \citenamefont {Bartschat},\ and\ \citenamefont
  {Schneider}}]{Guan2011}%
  \BibitemOpen
  \bibfield  {author} {\bibinfo {author} {\bibfnamefont {X.}~\bibnamefont
  {Guan}}, \bibinfo {author} {\bibfnamefont {E.~B.}\ \bibnamefont {Secor}},
  \bibinfo {author} {\bibfnamefont {K.}~\bibnamefont {Bartschat}}, \ and\
  \bibinfo {author} {\bibfnamefont {B.~I.}\ \bibnamefont {Schneider}},\ }\href
  {\doibase 10.1103/PhysRevA.84.033420} {\bibfield  {journal} {\bibinfo
  {journal} {Phys. Rev. A}\ }\textbf {\bibinfo {volume} {84}},\ \bibinfo
  {pages} {033420} (\bibinfo {year} {2011})}\BibitemShut {NoStop}%
\bibitem [{\citenamefont {{L. R. Moore, M. A. Lysaght, L. A. A. Nikolopoulos,
  J. S. Parker, and H. W. van der Hart and K. T.
  Taylor}}(2011)}]{Moore2EtAl2011}%
  \BibitemOpen
  \bibfield  {author} {\bibinfo {author} {\bibnamefont {{L. R. Moore, M. A.
  Lysaght, L. A. A. Nikolopoulos, J. S. Parker, and H. W. van der Hart and K.
  T. Taylor}}},\ }\href@noop {} {\bibfield  {journal} {\bibinfo  {journal} {J.
  Mod. Opt.}\ }\textbf {\bibinfo {volume} {58}},\ \bibinfo {pages} {1132}
  (\bibinfo {year} {2011})}\BibitemShut {NoStop}%
\bibitem [{\citenamefont {Dundas}(2012)}]{Dundas2012}%
  \BibitemOpen
  \bibfield  {author} {\bibinfo {author} {\bibfnamefont {D.}~\bibnamefont
  {Dundas}},\ }\href@noop {} {\bibfield  {journal} {\bibinfo  {journal} {J.
  Chem. Phys.}\ }\textbf {\bibinfo {volume} {136}},\ \bibinfo {pages} {194303}
  (\bibinfo {year} {2012})}\BibitemShut {NoStop}%
\bibitem [{\citenamefont {Carette}\ \emph {et~al.}(2013)\citenamefont
  {Carette}, \citenamefont {Dahlstr\"om}, \citenamefont {Argenti},\ and\
  \citenamefont {Lindroth}}]{argenti:2013}%
  \BibitemOpen
  \bibfield  {author} {\bibinfo {author} {\bibfnamefont {T.}~\bibnamefont
  {Carette}}, \bibinfo {author} {\bibfnamefont {J.~M.}\ \bibnamefont
  {Dahlstr\"om}}, \bibinfo {author} {\bibfnamefont {L.}~\bibnamefont
  {Argenti}}, \ and\ \bibinfo {author} {\bibfnamefont {E.}~\bibnamefont
  {Lindroth}},\ }\href {\doibase 10.1103/PhysRevA.87.023420} {\bibfield
  {journal} {\bibinfo  {journal} {Phys. Rev. A}\ }\textbf {\bibinfo {volume}
  {87}},\ \bibinfo {pages} {023420} (\bibinfo {year} {2013})}\BibitemShut
  {NoStop}%
\bibitem [{\citenamefont {Feist}\ \emph {et~al.}(2014)\citenamefont {Feist},
  \citenamefont {Zatsarinny}, \citenamefont {Nagele}, \citenamefont {Pazourek},
  \citenamefont {Burgd\"orfer}, \citenamefont {Guan}, \citenamefont
  {Bartschat},\ and\ \citenamefont {Schneider}}]{feist:2014}%
  \BibitemOpen
  \bibfield  {author} {\bibinfo {author} {\bibfnamefont {J.}~\bibnamefont
  {Feist}}, \bibinfo {author} {\bibfnamefont {O.}~\bibnamefont {Zatsarinny}},
  \bibinfo {author} {\bibfnamefont {S.}~\bibnamefont {Nagele}}, \bibinfo
  {author} {\bibfnamefont {R.}~\bibnamefont {Pazourek}}, \bibinfo {author}
  {\bibfnamefont {J.}~\bibnamefont {Burgd\"orfer}}, \bibinfo {author}
  {\bibfnamefont {X.}~\bibnamefont {Guan}}, \bibinfo {author} {\bibfnamefont
  {K.}~\bibnamefont {Bartschat}}, \ and\ \bibinfo {author} {\bibfnamefont
  {B.~I.}\ \bibnamefont {Schneider}},\ }\href {\doibase
  10.1103/PhysRevA.89.033417} {\bibfield  {journal} {\bibinfo  {journal} {Phys.
  Rev. A}\ }\textbf {\bibinfo {volume} {89}},\ \bibinfo {pages} {033417}
  (\bibinfo {year} {2014})}\BibitemShut {NoStop}%
\bibitem [{\citenamefont {Catoire}\ \emph {et~al.}(2014)\citenamefont
  {Catoire}, \citenamefont {Silva}, \citenamefont {Rivi\`{e}re}, \citenamefont
  {Bachau},\ and\ \citenamefont {Mart\'{\i}n}}]{Catoire2014}%
  \BibitemOpen
  \bibfield  {author} {\bibinfo {author} {\bibfnamefont {F.}~\bibnamefont
  {Catoire}}, \bibinfo {author} {\bibfnamefont {R.~E.~F.}\ \bibnamefont
  {Silva}}, \bibinfo {author} {\bibfnamefont {P.}~\bibnamefont {Rivi\`{e}re}},
  \bibinfo {author} {\bibfnamefont {H.}~\bibnamefont {Bachau}}, \ and\ \bibinfo
  {author} {\bibfnamefont {F.}~\bibnamefont {Mart\'{\i}n}},\ }\href {\doibase
  10.1103/PhysRevA.89.023415} {\bibfield  {journal} {\bibinfo  {journal} {Phys.
  Rev. A}\ }\textbf {\bibinfo {volume} {89}},\ \bibinfo {pages} {023415}
  (\bibinfo {year} {2014})}\BibitemShut {NoStop}%
\bibitem [{\citenamefont {Muller}(1999)}]{Muller1999a}%
  \BibitemOpen
  \bibfield  {author} {\bibinfo {author} {\bibfnamefont {H.}~\bibnamefont
  {Muller}},\ }\href {\doibase 10.1103/PhysRevA.60.1341} {\bibfield  {journal}
  {\bibinfo  {journal} {Phys. Rev. A}\ }\textbf {\bibinfo {volume} {60}},\
  \bibinfo {pages} {1341} (\bibinfo {year} {1999})}\BibitemShut {NoStop}%
\bibitem [{\citenamefont {Awasthi}\ \emph {et~al.}(2008)\citenamefont
  {Awasthi}, \citenamefont {Vanne}, \citenamefont {Saenz}, \citenamefont
  {Castro},\ and\ \citenamefont {Decleva}}]{Awasthi2008}%
  \BibitemOpen
  \bibfield  {author} {\bibinfo {author} {\bibfnamefont {M.}~\bibnamefont
  {Awasthi}}, \bibinfo {author} {\bibfnamefont {Y.~V.}\ \bibnamefont {Vanne}},
  \bibinfo {author} {\bibfnamefont {A.}~\bibnamefont {Saenz}}, \bibinfo
  {author} {\bibfnamefont {A.}~\bibnamefont {Castro}}, \ and\ \bibinfo {author}
  {\bibfnamefont {P.}~\bibnamefont {Decleva}},\ }\href {\doibase
  10.1103/PhysRevA.77.063403} {\bibfield  {journal} {\bibinfo  {journal} {Phys.
  Rev. A}\ }\textbf {\bibinfo {volume} {77}},\ \bibinfo {pages} {063403}
  (\bibinfo {year} {2008})}\BibitemShut {NoStop}%
\bibitem [{\citenamefont {Burke}\ \emph {et~al.}(1971)\citenamefont {Burke},
  \citenamefont {Hibbert},\ and\ \citenamefont {Robb}}]{Burke1971}%
  \BibitemOpen
  \bibfield  {author} {\bibinfo {author} {\bibfnamefont {P.~G.}\ \bibnamefont
  {Burke}}, \bibinfo {author} {\bibfnamefont {A.}~\bibnamefont {Hibbert}}, \
  and\ \bibinfo {author} {\bibfnamefont {W.~D.}\ \bibnamefont {Robb}},\
  }\href@noop {} {\bibfield  {journal} {\bibinfo  {journal} {J. Phys. B}\
  }\textbf {\bibinfo {volume} {4}},\ \bibinfo {pages} {153} (\bibinfo {year}
  {1971})}\BibitemShut {NoStop}%
\bibitem [{\citenamefont {Burke}(2011{\natexlab{a}})}]{Burke2011}%
  \BibitemOpen
  \bibfield  {author} {\bibinfo {author} {\bibfnamefont {P.~G.}\ \bibnamefont
  {Burke}},\ }\href@noop {} {\emph {\bibinfo {title} {{R-Matrix Theory of
  Atomic Collisions}}}}\ (\bibinfo  {publisher} {Springer Ser. At. Opt. Plasma
  Phys.},\ \bibinfo {year} {2011})\ p.\ \bibinfo {pages} {745}\BibitemShut
  {NoStop}%
\bibitem [{\citenamefont {Madden}\ \emph {et~al.}(2011)\citenamefont {Madden},
  \citenamefont {Tennyson},\ and\ \citenamefont {Zhang}}]{Madden2011}%
  \BibitemOpen
  \bibfield  {author} {\bibinfo {author} {\bibfnamefont {D.}~\bibnamefont
  {Madden}}, \bibinfo {author} {\bibfnamefont {J.}~\bibnamefont {Tennyson}}, \
  and\ \bibinfo {author} {\bibfnamefont {R.}~\bibnamefont {Zhang}},\ }\href
  {\doibase 10.1088/1742-6596/300/1/012017} {\bibfield  {journal} {\bibinfo
  {journal} {J. Phys. Conf. Ser.}\ }\textbf {\bibinfo {volume} {300}},\
  \bibinfo {pages} {012017} (\bibinfo {year} {2011})}\BibitemShut {NoStop}%
\bibitem [{\citenamefont {Burke}\ and\ \citenamefont
  {Robb}(1976)}]{Burke1975a}%
  \BibitemOpen
  \bibfield  {author} {\bibinfo {author} {\bibfnamefont {P.~G.}\ \bibnamefont
  {Burke}}\ and\ \bibinfo {author} {\bibfnamefont {W.~D.}\ \bibnamefont
  {Robb}},\ }\bibfield  {booktitle} {\emph {\bibinfo {booktitle} {Adv. At. Mol.
  Phys.}},\ }\href@noop {} {\ \textbf {\bibinfo {volume} {11}},\ \bibinfo
  {pages} {143} (\bibinfo {year} {1976})}\BibitemShut {NoStop}%
\bibitem [{\citenamefont {Burke}\ and\ \citenamefont
  {Taylor}(1975)}]{Burke1975b}%
  \BibitemOpen
  \bibfield  {author} {\bibinfo {author} {\bibfnamefont {P.~G.}\ \bibnamefont
  {Burke}}\ and\ \bibinfo {author} {\bibfnamefont {K.~T.}\ \bibnamefont
  {Taylor}},\ }\href@noop {} {\bibfield  {journal} {\bibinfo  {journal} {J.
  Phys. B}\ }\textbf {\bibinfo {volume} {8}},\ \bibinfo {pages} {2620}
  (\bibinfo {year} {1975})}\BibitemShut {NoStop}%
\bibitem [{\citenamefont {Burke}\ and\ \citenamefont
  {Burke}(1997)}]{burke:1997}%
  \BibitemOpen
  \bibfield  {author} {\bibinfo {author} {\bibfnamefont {P.~G.}\ \bibnamefont
  {Burke}}\ and\ \bibinfo {author} {\bibfnamefont {V.~M.}\ \bibnamefont
  {Burke}},\ }\href@noop {} {\bibfield  {journal} {\bibinfo  {journal} {J.
  Phys. B}\ }\textbf {\bibinfo {volume} {30}},\ \bibinfo {pages} {L383}
  (\bibinfo {year} {1997})}\BibitemShut {NoStop}%
\bibitem [{\citenamefont {van~der Hart}\ \emph {et~al.}(2007)\citenamefont
  {van~der Hart}, \citenamefont {Lysaght},\ and\ \citenamefont
  {Burke}}]{Hart2007}%
  \BibitemOpen
  \bibfield  {author} {\bibinfo {author} {\bibfnamefont {H.~W.}\ \bibnamefont
  {van~der Hart}}, \bibinfo {author} {\bibfnamefont {M.~A.}\ \bibnamefont
  {Lysaght}}, \ and\ \bibinfo {author} {\bibfnamefont {P.~G.}\ \bibnamefont
  {Burke}},\ }\href@noop {} {\bibfield  {journal} {\bibinfo  {journal} {Phys.
  Rev. A}\ }\textbf {\bibinfo {volume} {76}},\ \bibinfo {pages} {43405}
  (\bibinfo {year} {2007})}\BibitemShut {NoStop}%
\bibitem [{\citenamefont {Guan}\ \emph {et~al.}(2008)\citenamefont {Guan},
  \citenamefont {Noble}, \citenamefont {Zatsarinny}, \citenamefont
  {Bartschat},\ and\ \citenamefont {Schneider}}]{guan:2008}%
  \BibitemOpen
  \bibfield  {author} {\bibinfo {author} {\bibfnamefont {X.}~\bibnamefont
  {Guan}}, \bibinfo {author} {\bibfnamefont {C.~J.}\ \bibnamefont {Noble}},
  \bibinfo {author} {\bibfnamefont {O.}~\bibnamefont {Zatsarinny}}, \bibinfo
  {author} {\bibfnamefont {K.}~\bibnamefont {Bartschat}}, \ and\ \bibinfo
  {author} {\bibfnamefont {B.~I.}\ \bibnamefont {Schneider}},\ }\href {\doibase
  10.1103/PhysRevA.78.053402} {\bibfield  {journal} {\bibinfo  {journal} {Phys.
  Rev. A}\ }\textbf {\bibinfo {volume} {78}},\ \bibinfo {pages} {053402}
  (\bibinfo {year} {2008})}\BibitemShut {NoStop}%
\bibitem [{\citenamefont {Nikolopoulos}\ \emph {et~al.}(2008)\citenamefont
  {Nikolopoulos}, \citenamefont {Parker},\ and\ \citenamefont
  {Taylor}}]{Nikolopoulos2008}%
  \BibitemOpen
  \bibfield  {author} {\bibinfo {author} {\bibfnamefont {L.~A.~A.}\
  \bibnamefont {Nikolopoulos}}, \bibinfo {author} {\bibfnamefont {J.~S.}\
  \bibnamefont {Parker}}, \ and\ \bibinfo {author} {\bibfnamefont {K.~T.}\
  \bibnamefont {Taylor}},\ }\href {\doibase 10.1103/PhysRevA.78.063420}
  {\bibfield  {journal} {\bibinfo  {journal} {Phys. Rev. A}\ }\textbf {\bibinfo
  {volume} {78}},\ \bibinfo {pages} {063420} (\bibinfo {year}
  {2008})}\BibitemShut {NoStop}%
\bibitem [{\citenamefont {Lysaght}\ \emph {et~al.}(2011)\citenamefont
  {Lysaght}, \citenamefont {Moore}, \citenamefont {Nikolopoulos}, \citenamefont
  {Parker}, \citenamefont {Hart},\ and\ \citenamefont
  {Taylor}}]{LysaghtEtAl2011}%
  \BibitemOpen
  \bibfield  {author} {\bibinfo {author} {\bibfnamefont {M.~A.}\ \bibnamefont
  {Lysaght}}, \bibinfo {author} {\bibfnamefont {L.~R.}\ \bibnamefont {Moore}},
  \bibinfo {author} {\bibfnamefont {L.~A.~A.}\ \bibnamefont {Nikolopoulos}},
  \bibinfo {author} {\bibfnamefont {J.~S.}\ \bibnamefont {Parker}}, \bibinfo
  {author} {\bibfnamefont {H.~W.}\ \bibnamefont {Hart}}, \ and\ \bibinfo
  {author} {\bibfnamefont {K.~T.}\ \bibnamefont {Taylor}},\ }in\ \href@noop {}
  {\emph {\bibinfo {booktitle} {Quantum Dyn. Imaging}}},\ \bibinfo {series and
  number} {CRM Series in Mathematical Physics},\ \bibinfo {editor} {edited by\
  \bibinfo {editor} {\bibnamefont {{Andr\'{e} D. Bandrauk and Misha Ivanov}}}}\
  (\bibinfo  {publisher} {Springer New York},\ \bibinfo {year} {2011})\ pp.\
  \bibinfo {pages} {107--134}\BibitemShut {NoStop}%
\bibitem [{\citenamefont {Burke}\ \emph {et~al.}(1991)\citenamefont {Burke},
  \citenamefont {Francken},\ and\ \citenamefont {Joachain}}]{Burke1991}%
  \BibitemOpen
  \bibfield  {author} {\bibinfo {author} {\bibfnamefont {P.~G.}\ \bibnamefont
  {Burke}}, \bibinfo {author} {\bibfnamefont {P.}~\bibnamefont {Francken}}, \
  and\ \bibinfo {author} {\bibfnamefont {C.~J.}\ \bibnamefont {Joachain}},\
  }\href@noop {} {\bibfield  {journal} {\bibinfo  {journal} {J. Phys. B}\
  }\textbf {\bibinfo {volume} {24}},\ \bibinfo {pages} {751} (\bibinfo {year}
  {1991})}\BibitemShut {NoStop}%
\bibitem [{\citenamefont {Lysaght}\ \emph {et~al.}(2008)\citenamefont
  {Lysaght}, \citenamefont {Burke},\ and\ \citenamefont {van~der
  Hart}}]{lysaght:2008}%
  \BibitemOpen
  \bibfield  {author} {\bibinfo {author} {\bibfnamefont {M.~A.}\ \bibnamefont
  {Lysaght}}, \bibinfo {author} {\bibfnamefont {P.~G.}\ \bibnamefont {Burke}},
  \ and\ \bibinfo {author} {\bibfnamefont {H.~W.}\ \bibnamefont {van~der
  Hart}},\ }\href {\doibase 10.1103/PhysRevLett.101.253001} {\bibfield
  {journal} {\bibinfo  {journal} {Phys. Rev. Lett.}\ }\textbf {\bibinfo
  {volume} {101}},\ \bibinfo {pages} {253001} (\bibinfo {year}
  {2008})}\BibitemShut {NoStop}%
\bibitem [{\citenamefont {Hutchinson}\ \emph {et~al.}(2013)\citenamefont
  {Hutchinson}, \citenamefont {Lysaght},\ and\ \citenamefont {van~der
  Hart}}]{lysaght:2013}%
  \BibitemOpen
  \bibfield  {author} {\bibinfo {author} {\bibfnamefont {S.}~\bibnamefont
  {Hutchinson}}, \bibinfo {author} {\bibfnamefont {M.~A.}\ \bibnamefont
  {Lysaght}}, \ and\ \bibinfo {author} {\bibfnamefont {H.~W.}\ \bibnamefont
  {van~der Hart}},\ }\href {\doibase 10.1103/PhysRevA.88.023424} {\bibfield
  {journal} {\bibinfo  {journal} {Phys. Rev. A}\ }\textbf {\bibinfo {volume}
  {88}},\ \bibinfo {pages} {023424} (\bibinfo {year} {2013})}\BibitemShut
  {NoStop}%
\bibitem [{\citenamefont {Berrington}\ \emph {et~al.}(1995)\citenamefont
  {Berrington}, \citenamefont {Eissner},\ and\ \citenamefont
  {Norrington}}]{rmatrx1:1995}%
  \BibitemOpen
  \bibfield  {author} {\bibinfo {author} {\bibfnamefont {K.~A.}\ \bibnamefont
  {Berrington}}, \bibinfo {author} {\bibfnamefont {W.~B.}\ \bibnamefont
  {Eissner}}, \ and\ \bibinfo {author} {\bibfnamefont {P.~H.}\ \bibnamefont
  {Norrington}},\ }\href@noop {} {\bibfield  {journal} {\bibinfo  {journal}
  {Comput. Phys. Commun.}\ }\textbf {\bibinfo {volume} {92}},\ \bibinfo {pages}
  {290} (\bibinfo {year} {1995})}\BibitemShut {NoStop}%
\bibitem [{\citenamefont {Burke}(2011{\natexlab{b}})}]{burke:2011}%
  \BibitemOpen
  \bibfield  {author} {\bibinfo {author} {\bibfnamefont {P.~G.}\ \bibnamefont
  {Burke}},\ }\href@noop {} {\emph {\bibinfo {title} {{R-Matrix theory of
  Atomic Collisions}}}}\ (\bibinfo  {publisher} {Springer Series on Atomic,
  Optical and Plasma Physics},\ \bibinfo {year} {2011})\ p.\ \bibinfo {pages}
  {745}\BibitemShut {NoStop}%
\bibitem [{\citenamefont {Zatsarinny}(2006)}]{bsr:2006}%
  \BibitemOpen
  \bibfield  {author} {\bibinfo {author} {\bibfnamefont {O.}~\bibnamefont
  {Zatsarinny}},\ }\href {\doibase http://dx.doi.org/10.1016/j.cpc.2005.10.006}
  {\bibfield  {journal} {\bibinfo  {journal} {Comput. Phys. Commun.}\ }\textbf
  {\bibinfo {volume} {174}},\ \bibinfo {pages} {273} (\bibinfo {year}
  {2006})}\BibitemShut {NoStop}%
\bibitem [{\citenamefont {Guan}\ \emph {et~al.}(2009)\citenamefont {Guan},
  \citenamefont {Noble}, \citenamefont {Zatsarinny}, \citenamefont
  {Bartschat},\ and\ \citenamefont {Schneider}}]{guan2009}%
  \BibitemOpen
  \bibfield  {author} {\bibinfo {author} {\bibfnamefont {X.}~\bibnamefont
  {Guan}}, \bibinfo {author} {\bibfnamefont {C.~J.}\ \bibnamefont {Noble}},
  \bibinfo {author} {\bibfnamefont {O.}~\bibnamefont {Zatsarinny}}, \bibinfo
  {author} {\bibfnamefont {K.}~\bibnamefont {Bartschat}}, \ and\ \bibinfo
  {author} {\bibfnamefont {B.~I.}\ \bibnamefont {Schneider}},\ }\href@noop {}
  {\bibfield  {journal} {\bibinfo  {journal} {Comput. Phys. Commun.}\ }\textbf
  {\bibinfo {volume} {180}},\ \bibinfo {pages} {2401} (\bibinfo {year}
  {2009})}\BibitemShut {NoStop}%
\bibitem [{\citenamefont {Wragg}\ \emph {et~al.}(2015)\citenamefont {Wragg},
  \citenamefont {Parker},\ and\ \citenamefont {van~der
  Hart}}]{vanderhart:2015}%
  \BibitemOpen
  \bibfield  {author} {\bibinfo {author} {\bibfnamefont {J.}~\bibnamefont
  {Wragg}}, \bibinfo {author} {\bibfnamefont {J.~S.}\ \bibnamefont {Parker}}, \
  and\ \bibinfo {author} {\bibfnamefont {H.~W.}\ \bibnamefont {van~der Hart}},\
  }\href {\doibase 10.1103/PhysRevA.92.022504} {\bibfield  {journal} {\bibinfo
  {journal} {Phys. Rev. A}\ }\textbf {\bibinfo {volume} {92}},\ \bibinfo
  {pages} {022504} (\bibinfo {year} {2015})}\BibitemShut {NoStop}%
\bibitem [{\citenamefont {Tao}\ and\ \citenamefont {Scrinzi}(2012)}]{Tao2012}%
  \BibitemOpen
  \bibfield  {author} {\bibinfo {author} {\bibfnamefont {L.}~\bibnamefont
  {Tao}}\ and\ \bibinfo {author} {\bibfnamefont {A.}~\bibnamefont {Scrinzi}},\
  }\href@noop {} {\bibfield  {journal} {\bibinfo  {journal} {New J. Phys.}\
  }\textbf {\bibinfo {volume} {14}},\ \bibinfo {pages} {013021} (\bibinfo
  {year} {2012})}\BibitemShut {NoStop}%
\bibitem [{\citenamefont {Scrinzi}(2012)}]{Scrinzi2012}%
  \BibitemOpen
  \bibfield  {author} {\bibinfo {author} {\bibfnamefont {A.}~\bibnamefont
  {Scrinzi}},\ }\href@noop {} {\bibfield  {journal} {\bibinfo  {journal} {New
  J. Phys.}\ }\textbf {\bibinfo {volume} {14}},\ \bibinfo {pages} {085008}
  (\bibinfo {year} {2012})}\BibitemShut {NoStop}%
\bibitem [{\citenamefont {Torlina}\ and\ \citenamefont
  {Smirnova}(2012)}]{Torlina2012a}%
  \BibitemOpen
  \bibfield  {author} {\bibinfo {author} {\bibfnamefont {L.}~\bibnamefont
  {Torlina}}\ and\ \bibinfo {author} {\bibfnamefont {O.}~\bibnamefont
  {Smirnova}},\ }\href@noop {} {\bibfield  {journal} {\bibinfo  {journal}
  {Phys. Rev. A}\ }\textbf {\bibinfo {volume} {86}},\ \bibinfo {pages} {043408}
  (\bibinfo {year} {2012})}\BibitemShut {NoStop}%
\bibitem [{\citenamefont {Torlina}\ \emph {et~al.}(2012)\citenamefont
  {Torlina}, \citenamefont {Ivanov}, \citenamefont {Walters},\ and\
  \citenamefont {Smirnova}}]{Torlina2012b}%
  \BibitemOpen
  \bibfield  {author} {\bibinfo {author} {\bibfnamefont {L.}~\bibnamefont
  {Torlina}}, \bibinfo {author} {\bibfnamefont {M.}~\bibnamefont {Ivanov}},
  \bibinfo {author} {\bibfnamefont {Z.~B.}\ \bibnamefont {Walters}}, \ and\
  \bibinfo {author} {\bibfnamefont {O.}~\bibnamefont {Smirnova}},\ }\href@noop
  {} {\bibfield  {journal} {\bibinfo  {journal} {Phys. Rev. A}\ }\textbf
  {\bibinfo {volume} {86}},\ \bibinfo {pages} {043409} (\bibinfo {year}
  {2012})}\BibitemShut {NoStop}%
\bibitem [{\citenamefont {Torlina}\ \emph {et~al.}(2014)\citenamefont
  {Torlina}, \citenamefont {Morales}, \citenamefont {Muller},\ and\
  \citenamefont {Smirnova}}]{Torlina2014}%
  \BibitemOpen
  \bibfield  {author} {\bibinfo {author} {\bibfnamefont {L.}~\bibnamefont
  {Torlina}}, \bibinfo {author} {\bibfnamefont {F.}~\bibnamefont {Morales}},
  \bibinfo {author} {\bibfnamefont {H.}~\bibnamefont {Muller}}, \ and\ \bibinfo
  {author} {\bibfnamefont {O.}~\bibnamefont {Smirnova}},\ }\href@noop {}
  {\bibfield  {journal} {\bibinfo  {journal} {J. Phys. B}\ }\textbf {\bibinfo
  {volume} {47}},\ \bibinfo {pages} {204021} (\bibinfo {year}
  {2014})}\BibitemShut {NoStop}%
\bibitem [{\citenamefont {Cormier}\ and\ \citenamefont
  {Lambropoulos}(1996)}]{Cormier1996}%
  \BibitemOpen
  \bibfield  {author} {\bibinfo {author} {\bibfnamefont {E.}~\bibnamefont
  {Cormier}}\ and\ \bibinfo {author} {\bibfnamefont {P.}~\bibnamefont
  {Lambropoulos}},\ }\href {http://stacks.iop.org/0953-4075/29/i=9/a=013}
  {\bibfield  {journal} {\bibinfo  {journal} {J. Phys. B}\ }\textbf {\bibinfo
  {volume} {77}},\ \bibinfo {pages} {1667} (\bibinfo {year}
  {1996})}\BibitemShut {NoStop}%
\bibitem [{\citenamefont {{\'{O} Broin}}\ and\ \citenamefont
  {Nikolopoulos}(2012)}]{OBroinNikolopoulos2012}%
  \BibitemOpen
  \bibfield  {author} {\bibinfo {author} {\bibfnamefont {C.}~\bibnamefont
  {{\'{O} Broin}}}\ and\ \bibinfo {author} {\bibfnamefont {L.~A.~A.}\
  \bibnamefont {Nikolopoulos}},\ }\href {\doibase 10.1016/j.cpc.2012.05.009}
  {\bibfield  {journal} {\bibinfo  {journal} {Comput. Phys. Commun.}\ }\textbf
  {\bibinfo {volume} {183}},\ \bibinfo {pages} {2071} (\bibinfo {year}
  {2012})},\ \Eprint {http://arxiv.org/abs/1201.6062} {1201.6062} \BibitemShut
  {NoStop}%
\bibitem [{\citenamefont {{\'{O} Broin}}\ and\ \citenamefont
  {Nikolopoulos}(2014)}]{OBroin2014}%
  \BibitemOpen
  \bibfield  {author} {\bibinfo {author} {\bibfnamefont {C.}~\bibnamefont
  {{\'{O} Broin}}}\ and\ \bibinfo {author} {\bibfnamefont {L.~A.~A.}\
  \bibnamefont {Nikolopoulos}},\ }\href@noop {} {\bibfield  {journal} {\bibinfo
   {journal} {Comput. Phys. Commun.}\ }\textbf {\bibinfo {volume} {185}},\
  \bibinfo {pages} {1791} (\bibinfo {year} {2014})}\BibitemShut {NoStop}%
\bibitem [{\citenamefont {Bransden}\ and\ \citenamefont
  {Joachain}(2003)}]{Bransden2003}%
  \BibitemOpen
  \bibfield  {author} {\bibinfo {author} {\bibfnamefont {B.~H.}\ \bibnamefont
  {Bransden}}\ and\ \bibinfo {author} {\bibfnamefont {C.~J.}\ \bibnamefont
  {Joachain}},\ }\href@noop {} {\emph {\bibinfo {title} {Physics of Atoms and
  Molecules}}}\ (\bibinfo  {publisher} {Pearson Education India},\ \bibinfo
  {year} {2003})\BibitemShut {NoStop}%
\bibitem [{\citenamefont {Dundas}\ \emph {et~al.}(2000)\citenamefont {Dundas},
  \citenamefont {McCann}, \citenamefont {Parker},\ and\ \citenamefont
  {Taylor}}]{Dundas2000}%
  \BibitemOpen
  \bibfield  {author} {\bibinfo {author} {\bibfnamefont {D.}~\bibnamefont
  {Dundas}}, \bibinfo {author} {\bibfnamefont {J.~F.}\ \bibnamefont {McCann}},
  \bibinfo {author} {\bibfnamefont {J.~S.}\ \bibnamefont {Parker}}, \ and\
  \bibinfo {author} {\bibfnamefont {K.~T.}\ \bibnamefont {Taylor}},\ }\href
  {http://stacks.iop.org/0953-4075/33/i=17/a=308} {\bibfield  {journal}
  {\bibinfo  {journal} {J. Phys. B}\ }\textbf {\bibinfo {volume} {33}},\
  \bibinfo {pages} {3261} (\bibinfo {year} {2000})}\BibitemShut {NoStop}%
\bibitem [{\citenamefont {Mart\'{\i}n}(1999)}]{Martin1999}%
  \BibitemOpen
  \bibfield  {author} {\bibinfo {author} {\bibfnamefont {F.}~\bibnamefont
  {Mart\'{\i}n}},\ }\href@noop {} {\bibfield  {journal} {\bibinfo  {journal}
  {J. Phys. B}\ }\textbf {\bibinfo {volume} {32}},\ \bibinfo {pages} {R197}
  (\bibinfo {year} {1999})}\BibitemShut {NoStop}%
\bibitem [{\citenamefont {Bachau}\ \emph {et~al.}(2001)\citenamefont {Bachau},
  \citenamefont {Cormier}, \citenamefont {Decleva}, \citenamefont {Hansen},\
  and\ \citenamefont {Mart\'in}}]{Bachau2001}%
  \BibitemOpen
  \bibfield  {author} {\bibinfo {author} {\bibfnamefont {H.}~\bibnamefont
  {Bachau}}, \bibinfo {author} {\bibfnamefont {E.}~\bibnamefont {Cormier}},
  \bibinfo {author} {\bibfnamefont {P.}~\bibnamefont {Decleva}}, \bibinfo
  {author} {\bibfnamefont {J.~E.}\ \bibnamefont {Hansen}}, \ and\ \bibinfo
  {author} {\bibfnamefont {F.}~\bibnamefont {Mart\'in}},\ }\href
  {http://stacks.iop.org/0034-4885/64/i=12/a=205} {\bibfield  {journal}
  {\bibinfo  {journal} {Reports Prog. Phys.}\ }\textbf {\bibinfo {volume}
  {64}},\ \bibinfo {pages} {1815} (\bibinfo {year} {2001})}\BibitemShut
  {NoStop}%
\bibitem [{\citenamefont {Barmaki}\ \emph {et~al.}(2003)\citenamefont
  {Barmaki}, \citenamefont {Laulan}, \citenamefont {Bachau},\ and\
  \citenamefont {Ghalim}}]{Barmaki2003}%
  \BibitemOpen
  \bibfield  {author} {\bibinfo {author} {\bibfnamefont {S.}~\bibnamefont
  {Barmaki}}, \bibinfo {author} {\bibfnamefont {S.}~\bibnamefont {Laulan}},
  \bibinfo {author} {\bibfnamefont {H.}~\bibnamefont {Bachau}}, \ and\ \bibinfo
  {author} {\bibfnamefont {M.}~\bibnamefont {Ghalim}},\ }\href {\doibase
  10.1088/0953-4075/36/5/303} {\bibfield  {journal} {\bibinfo  {journal} {J.
  Phys. B}\ }\textbf {\bibinfo {volume} {36}},\ \bibinfo {pages} {817}
  (\bibinfo {year} {2003})}\BibitemShut {NoStop}%
\bibitem [{\citenamefont {Bloch}(1957)}]{Bloch1957}%
  \BibitemOpen
  \bibfield  {author} {\bibinfo {author} {\bibfnamefont {C.}~\bibnamefont
  {Bloch}},\ }\href@noop {} {\bibfield  {journal} {\bibinfo  {journal} {Nucl.
  Phys.}\ }\textbf {\bibinfo {volume} {4}},\ \bibinfo {pages} {503} (\bibinfo
  {year} {1957})}\BibitemShut {NoStop}%
\bibitem [{\citenamefont {Cohen-Tannoudji}\ \emph {et~al.}(1977)\citenamefont
  {Cohen-Tannoudji}, \citenamefont {Diu},\ and\ \citenamefont
  {Lal\"{o}e}}]{QuantumMechanicsVol1Tannoudji}%
  \BibitemOpen
  \bibfield  {author} {\bibinfo {author} {\bibfnamefont {C.}~\bibnamefont
  {Cohen-Tannoudji}}, \bibinfo {author} {\bibfnamefont {B.}~\bibnamefont
  {Diu}}, \ and\ \bibinfo {author} {\bibfnamefont {F.}~\bibnamefont
  {Lal\"{o}e}},\ }\href@noop {} {\emph {\bibinfo {title} {{Quantum Mechanics:
  Volume 1}}}}\ (\bibinfo  {publisher} {Wiley-VCH Verlag GmbH},\ \bibinfo
  {year} {1977})\ p.\ \bibinfo {pages} {914}\BibitemShut {NoStop}%
\end{thebibliography}%
% \printglossaries

\begin{acronym}
\acro{AMO}{Atomic, Molecular and Optical Physics}
\acro{BS}{Bond Softening}
\acro{BO}{Born-Oppenheimer}
\acro{EM}{Electro-magnetic}
\acro{FFT}{Fast Fourier Transform}
\acro{RKF}{Runge-Kutta-Fehlberg}
\acro{FEL}{Free-Electron laser}
\acro{FWHM}{Full-Width at Half-Maximum}
\acro{CEP}{Carrier-Envelope Phase}
\acro{HHG}{High Harmonic Generation}
\acro{SAE}{Single Active Electron}
\acro{HPC}{High-Performance Computing}
\acro{LOPT}{Lowest (non-vanishing) Order Perturbation Theory}
\acro{ATI}{Above-Threshold Ionization}
\acro{ATD}{Above-Threshold Dissociation}
\acro{PES}{Photo-electron Spectrum}
\acro{TDDFT}{Time-dependent Density Functional Theory}
\acro{TD}{time-dependent}
\acro{TDRM}{Time Dependent R-Matrix}
\acro{RMT}{R-Matrix incorporating Time}
\acro{TDSE}{Time-dependent Schr\"odinger Equation}
\acro{TISE}{Time-independent Schr\"odinger Equation}
\acro{GPGPU}{General Purpose Computation/Computing on Graphical Processing Unit}
\acro{MPI}{Message Passing Interface}
\acro{FD}{Finite Difference}
\acro{PDE}{partial differential equation}
\acro{ODE}{ordinary differential equation}
\end{acronym}

\end{document}